\documentclass[
aps,
 amsmath,amssymb,
 reprint,%
superscriptaddress,
nofootinbib,
showkeys,
floatfix
]{revtex4-2}
\let\Gamma\varGamma
\let\Delta\varDelta
\let\Theta\varTheta
\let\Lambda\varLambda
\let\Xi\varXi
\let\Pi\varPi
\let\Sigma\varSigma
\let\Upsilon\varUpsilon
\let\Phi\varPhi
\let\Psi\varPsi
\let\Omega\varOmega

\usepackage{tikz}
\usetikzlibrary{positioning,shapes}

\usepackage{tikz,xcolor}%orcid
\definecolor{lime}{HTML}{A6CE39}
\DeclareRobustCommand{\orcidicon}{%
	\begin{tikzpicture}
	\draw[lime, fill=lime] (0,0) 
	circle [radius=0.16] 
	node[white] {{\fontfamily{qag}\selectfont \tiny ID}};
	\draw[white, fill=white] (-0.06475,0.095) 
	circle [radius=0.007];
	\end{tikzpicture}
	\hspace{-2mm}
}
\foreach \x in {A, ..., Z}{%
	\expandafter\xdef\csname orcid\x\endcsname{\noexpand\href{https://orcid.org/\csname orcidauthor\x\endcsname}{\noexpand\orcidicon}}
}

\usepackage[utf8]{inputenc}
\usepackage[T1]{fontenc}
\usepackage{lmodern}
\usepackage{etoolbox}
\usepackage{siunitx}

\usepackage{graphicx}
\usepackage{dcolumn}
\usepackage{array}
\usepackage{booktabs}
\usepackage{colortbl}
\usepackage{xcolor}
\usepackage{bm}
\usepackage[hidelinks,allcolors=blue!60!green]{hyperref}

\hypersetup{
    unicode=false,
    pdftoolbar=true,
    pdfmenubar=true,
    pdffitwindow=false,
    pdfstartview={XYZ null null 1.00},
    colorlinks=true,
    pdfnewwindow=true,
    colorlinks=true,
}
\begin{document}
%%%%%%%%%%%%%%%%%%%%%%%%%%%%%%%%%%%%%%%%%%%%%

%\preprint{preprint-text}

\title{Simulation of sympathetic cooling efficiency in a linear Paul trap\\driven by alternative waveforms}

\author{O. Sund\orcidA{}}
\email{sund@qute.uni-hannover.de}

\affiliation{Institut für Festkörperphysik, Leibniz Universität Hannover, Appelstr. 2, 30167 Hannover, Germany}

\author{A. W. Schell\orcidB{}}
\affiliation{Johannes Kepler Universität Linz, Altenberger Straße 69, 4040 Linz, Austria}
\affiliation{Physikalisch-Technische Bundesanstalt, Bundesallee 100, 38116 Braunschweig, Germany}
\affiliation{Institut für Festkörperphysik, Leibniz Universität Hannover, Appelstr. 2, 30167 Hannover, Germany}
\date{\today}
%%%%%%%%%%%%%%%%%%%%%%%%%%%%%%%%%%%%%%%%%%%%%
\begin{abstract}
Cooling of ions or other charged particles in electromagnetic traps is an essential tool to achieve control over their degrees of freedom on the quantum level. For many objects, there is no viable route for direct cooling, such as an accessible laser cooling transition. In such a case, the sympathetic cooling can be used, where a particle with such a direct route is used to cool down the other particle via Coulomb interaction. On the downside, this cooling process often is inefficient. Here, we numerically evaluate the sympathetic cooling performance in a quadrupole ion trap for different driving waveforms. We find that using different driving waveforms and optimized trap parameters the sympathetic cooling performance can be enhanced. These results will open up the way to achieve larger sympathetic cooling rates from which many techniques, such as aluminum ion clocks, might profit.
\end{abstract}
%%%%%%%%%%%%%%%%%%%%%%%%%%%%%%%%%%%%%%%%%%%%%
\keywords{Paul trap, Sympathetic cooling, RF waveform, Stability diagram}
\maketitle
%%%%%%%%%%%%%%%%%%%%%%%%%%%%%%%%%%%%%%%%%%%%%
\section{Introduction}
Trapping of charged particles is an important technique in many fields of
physics: trapped ions~\cite{Dehmelt1990} are one of the prime systems used for quantum computation~\cite{Cirac1995,Schmidt-Kaler2003,Bruzewicz2019,Friis2018} and can also be used in optical clocks~\cite{Chou2010,Brewer2019}. One way to hold charged particles in place and decouple them from the environment are Paul traps~\cite{Paul1953}, where a time-varying electric quadrupole trapping field is applied.

In order to control the motion of a trapped particle and prevent decoherence of the quantum state due to uncontrolled motion, cooling techniques such as Doppler cooling are commonly used~\cite{Neuhauser1980,Roos1999}. While laser cooling is highly efficient, some ions, such as Al$^{+}$ which is frequently used in quantum logic clocks~\cite{Chou2010}, do not possess easily accessible optical transitions suited for laser cooling.
Nevertheless such ions can be controlled by a process called sympathetic cooling~\cite{Larson1986,Wineland1995}. In the sympathetic cooling process of ions, a species that is technically challenging to laser cool is coupled with ions of a species that is more easy to cool. The coupling is mediated by the Coulomb force between the particles. Now, if one particle is cooled, the other particle will be also cooled as it will transfer excitations to the cooled particle.  

It is noteworthy that Paul traps are not limited to ions, but can also be used to trap
nano and micro particles as long as they are charged, which enables for spectroscopy
of systems decoupled from the environment, e.g., of aerosols~\cite{Arnold1984}, plasmonic particles~\cite{Schell2017,Berthelot2019}, or solid-state quantum emitters~\cite{Kuhlicke2014,Delord2017}. Furthermore, the method of sympathetic cooling can be extended to nanoparticles in Paul traps~\cite{Bykov2023}.

As trapping in a Paul trap is highly dependent on the charge to mass ratio of the trapped objects, such traps can function as mass spectrometers~\cite{Paul1953,March1997}.

In order to cool the center of mass motion of particles in a Paul trap it is not
only possible to apply the laser cooling techniques, but also feedback cooling, which has been demonstrated for ions~\cite{Bushev2006}, nanoparticles~\cite{Conangla2018}, and even two-dimensional graphene flakes~\cite{Nagornykh2015}. In feedback cooling a precise position measurement is used, e.g., by interferometry, to adjust the trapping voltage amplitude in a way to extract energy from the moving particle.

Usually, Paul traps are driven by sinusoidal waveforms. Here, we are going to investigate the effects that arise from changing the trapping waveform to other functions.

Simulations showed that simultaneous trapping of two particle species of different charge-to-mass ratio can be achieved by operating with a two-frequency driven Paul trap~\cite{Trypo2016}. This so-called dual radiofrequency (RF) technique has also been part of numerical studies in context of sympathetic cooling efficiency~\cite{Meng2021}. 

So far, studies on Paul traps driven by non-sinusoidal RF waveforms have been focused on the rectangular waveform primarily for mass-spectrometer applications: ranging from simulation and experimental studies on the stability diagram of the digital ion trap (DIT) at various duty cycles~\cite{Ding2006,Bandelow2013} to their use for mass-selective resonant ejection scans~\cite{Ding2002,Koizumi2009,Brancia2010} and modulation of trap frequency and voltage pairing, giving raise to multiple pass bands of mass selection~\cite{Brabeck2016}. Matrix calculations show different stability limits and pseudo-potential depths in quadrupole traps driven by rectangular, triangular and trapezoid waveforms, potentially allowing for improving the mass resolving power~\cite{Reilly2015,Brabeck2016B}. Furthermore, rectangular and triangular driven mass filters can improve frequency scanning~\cite{Xiong2012} and excite protein and cluster ions by changing the duty cycle, minimizing solvent clustering~\cite{Opacic2018}. Simulation based comparison of single ion dynamic in a quadrupole trap driven by sinusoidal, rectangular, triangular and sawtooth waveforms revealed differences in oscillation amplitudes~\cite{Aksakal2016} and temperature of a trapped ion ensemble~\cite{Aksakal2020}. 

While the digital waveform technology has established in mass-filter applications, only few studies on advantages in using alternative trap driving waveforms for sympathetic cooling of particles in a Paul trap are existent to date. In this work, we are especially interested in the comparison of the sympathetic cooling performance of the sinusoidal wave with its alternatives. 

Due to initially high temperatures and low Coulomb-coupling, sympathetic cooling of ions until reaching Coulomb-crystallization can take up to several hundreds milliseconds, compared to microseconds timescales of direct Doppler laser cooling~\cite{Phillips1989,Drewsen2003,Schiller2003,Wübenna2012}. This is a crucial limitation for atomic-clock cycle scans and preparing ions for quantum engineering. We therefore attempt in this work to numerically investigate the possibilities in reducing the sympathetic cooling times in a linear Paul trap by applying alternative driving waveforms.
%%%%%%%%%%%%%%%%%%%%%%%%%%%%%%%%%%%%%%%%%%%%%
\section{\label{sec:level2}Simulation Model}
%%%%%%%%%%%%%%%%%%%%%%%%%%%%%%%%%%%%%%%%%%%%%
The simulation model is based on numerically solving Newton's equations of motion for a classical two-ion system confined in a linear Paul trap driven by different waveforms and interacting via the Coulomb force under laser cooling. The system equations are solved for a random sample of initial conditions of both particles, different trap driving waveform and trap voltages. Throughout the system's evolution, the cooling dynamics is evaluated as described in the following Sec.~\ref{subsec:evaluation}.
\subsection{Sympathetic cooling model}
\label{subsec:simulation_model}
In this subsection, the basics of the linear Paul trap and the sympathetic cooling model are introduced.
\subsubsection{Trap equations}
\begin{figure}[t!]
\includegraphics[width=0.475\textwidth]{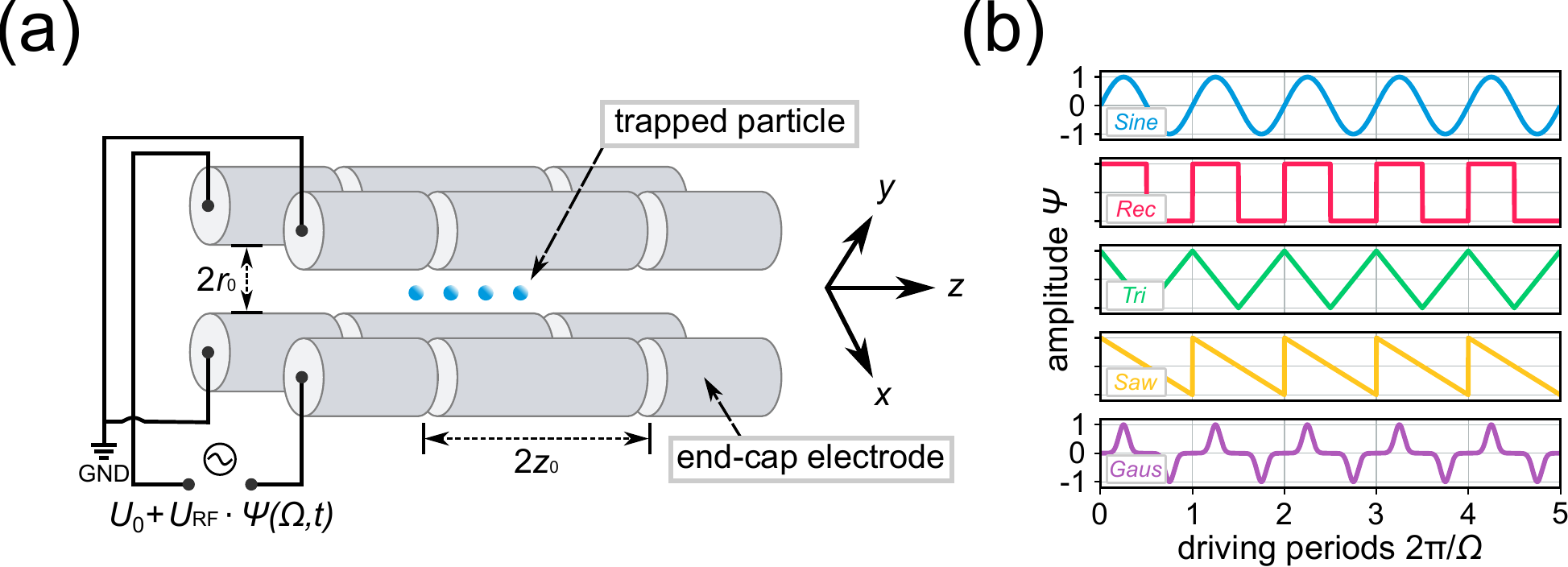}
\caption{\label{fig:trapdriv}(a) Sketch of a linear Paul trap. Particle confinement in the $x$- and $y$-direction is achieved by application of a time-varying voltage $U_0+U_{\text{RF}}\cdot\Psi(\Omega,t)$ to diagonally opposing rods. $z$-confinement is realized by e.g. additional end-cap electrodes with applied constant voltage $U_z$ on diagonal segments with respect to ground. Trapped particles then align along the axial trap axis. (b) Time-varying amplitude $\Psi(\Omega,t)$ of each trap driving waveform studied in this work.}
\label{fig:trap}
\end{figure}

In linear Paul traps, confinement of charged particles is realized via an rapidly oscillating electrical quadrupole field, generated by a four-electrode configuration as shown in Fig.~\ref{fig:trap}(a)~\cite{Wuerker1959,Paul1990}. Since the resulting ponderomotive restoring force only allows for a radial ($x$-$y$-plane) confinement, additional end-cap or segmentation electrodes are used to generate a static harmonic potential in the axial direction ($z$-axis), thereby allowing for a three dimensional trapping of the particles~\cite{Drewsen2000,Berkeland1998,Wineland1999}. 

With respect to the geometry shown in Fig.~\ref{fig:trap}(a), the time-dependent trap potential $V_{\text{trap}}$ is assumed to be\;\, approximated by
\begin{align}
    V_{\text{trap}}(x,y,z,t) =\, &\dfrac{1}{2}(U_0+U_{\text{RF}} \Psi(\Omega,t))\Bigl[1+\dfrac{x^2-y^2}{r_0^2}\Bigr] \nonumber\\
    &+ \dfrac{U_z}{z_0^2}\Bigl[z^2-\dfrac{1}{2}(x^2+y^2) \Bigr], 
    \label{eq:trap_potential}
\end{align}
where $U_{\text{RF}}$ and $U_0$ are applied RF, respectively constant offset trap voltage, generating the electric quadrupole field in the $x$-$y$-plane and $r_0$ is half the electrodes radial separation distance~\cite{Paul1990,Berkeland1998,Drewsen2000,Trypo2016,Wineland1999}. The axial potential is described by an ideal parabolic potential (second term) and its shape depends on axial trap voltage $U_z$ and distance from the trap center $z_0$. Note that the additional axial potential entails radial defocusing components such that the total trap potential $V_{\text{trap}}$ still satisfies the Laplace-equation, $\nabla^2V_{\text{trap}} = 0$.

For the sake of this study, we replaced the usually used sinusoidal term with an arbitrary periodic waveform $\Psi(\Omega,t)=\Psi(\Omega,t+2\pi/\Omega)$, where $\Omega=\,\textit{const.}$ is the fundamental driving frequency and $\vert\Psi(\Omega,t)\vert\leq 1\,\forall t$. The simulated driving waveforms (shown from top to bottom in Fig.~\ref{fig:trap}(b)) are: a sinusoidal wave (\textsl{Sine}), a rectangular (\textsl{Rec}) and a triangular (\textsl{Tri}) function with a duty cycle of 50\%, a sawtooth (\textsl{Saw}), and an example of a more exotic waveform with a Gaussian shape (\textsl{Gaus}). The resulting equations of motion $m\Ddot{\bm{x}}=-Q\nabla V_{\text{trap}}$, for a particle of mass $m$ and charge $Q$ at position $\bm{x} = (x,y,z)$, can be rewritten into generalized case of Hill differential equations~\cite{Hill1900,Hill2004} by the common substitution $\xi=\Omega t/2$:
\begin{subequations}
\begin{eqnarray}
    &\dfrac{d^2 x}{d\xi^2}+ \bigl(a+2q\Psi(\xi)-a_z\bigr)\cdot x(\xi) &= 0,\label{eq:hill_1}\\
    &\dfrac{d^2 y}{d\xi^2} - \bigl(a+2q\Psi(\xi)+a_z\bigr)\cdot y(\xi) &= 0,\label{eq:hill_2}\\
    &\dfrac{d^2 z}{d\xi^2} + 2a_z\cdot z(\xi) &=0.\label{eq:hill_dgl3}
\end{eqnarray}
\end{subequations}
This derivation is only applicable for non-chirping waveforms $\Omega\neq \Omega(t)$, since the chain-rule would imply additional, more complex terms.

In case of the traditional waveform $\Psi(\xi)=\cos(2\xi)$, Eqs.~\ref{eq:hill_1} and \ref{eq:hill_2} represent the well-known Mathieu differential  equations~\cite{Mathieu1868,Meixner1954,Abramowitz1988,Kovacic2018}. In this convention, as well as in our generalized case, dimensionless parameters are defined as $q=2QU_{\text{RF}}/(m\Omega^2r_0^2)$, $a=4QU_{0}/(m\Omega^2r_0^2)$, and $a_z =4Q U_z/(m\Omega^2z_0^2)$, fully describing the dynamic of a single trapped particle~\cite{Paul1990,Berkeland1998,Drewsen2000}. Furthermore, only certain combinations of $q$, $a$ and $a_z$ can generate a stable particle confinement, while others lead to an exponential growth of the particle's kinetic energy causing the particle to leave any finite trap. 

In the so-called adiabatic approximation, meaning sufficient high driving frequency $\Omega$ (such that $\vert q\vert\ll 0.4$ at the sinusoidal-driven trap), Eq.~\ref{eq:trap_potential} can be averaged to an effective pseudo-potential~\cite{Dehmelt1968,Major2005}. For the sinusoidal-driven trap, this potential $\Phi$ is given by~\cite{Major2005,Guan1994}
\begin{equation}
    \Phi_{\text{trap}}(x,y,z) = \tilde{D_r}(x^2+y^2)+2D_z z^2,
    \label{eq:pseudo_potential}
\end{equation}
for an offset voltage of $a=0$. Here, $\tilde{D_r}=D_r-D_z$ is the reduced radial well depth due to the defocusing effect of the axial field, $D_r=QU_{\text{RF}}^2/(4m\Omega^2r_0^4)=qU_{\text{RF}}/(8r_0^2)$ and $D_z=U_z/(2z_0^2)$. In the case  $a\neq 0$, the $x$- and $y$-well depths' respectively oscillation frequencies would become unequal. Using these equations, the maximum kinetic energy of a single trapped particle can be estimated.
\subsubsection{Sympathetic cooling model}
The classical treatment of a sympathetic cooling process involves the trapping force derived from Eq.~\ref{eq:trap_potential} and Coulomb forces between both ions~\cite{Schiller2003}. Laser cooling on one ion is modeled by a linear damping force proportional to the ion's velocity in all three dimensions~\cite{Zhang2007,Guggemos2015}.
The equations of motion of sympathetic cooled ion (noted as particle A) and coolant ion (particle B) can therefore be expressed as
\begin{subequations}
\begin{eqnarray}
m_{\text{A}}\ddot{\bm{x}}_{\text{A}}=e\bm{E}_{\text{trap}}+\dfrac{e^2}{4\pi \epsilon_0}\dfrac{(\bm{x}_{\text{A}}-\bm{x}_{\text{B}})}{R_{\text{AB}}^3} \, , \label{eq:eqm1} \\
    m_{\text{B}}\ddot{\bm{x}}_{\text{B}}=-\beta_{\text{c}} \dot{\bm{x}}_{\text{B}}+e\bm{E}_{\text{trap}}+\dfrac{e^2}{4\pi \epsilon_0}\dfrac{(\bm{x}_{\text{B}}-\bm{x}_{\text{A}})}{R_{\text{AB}}^3} \, ,
    \label{eq:eqm2}
\end{eqnarray}
\end{subequations}
where $R_{\text{AB}} = \vert \bm{x}_{\text{A}}-\bm{x}_{\text{B}}\vert $ is the ions' distance, $1e$ their charges ($e=1.6022\times 10^{-19}$ C), $m$ the masses, and $\bm{E}_{\text{trap}}=-\nabla V_{\text{trap}}$. $\beta_{\text{c}}$ is the damping coefficient of ion B due to the laser cooling, $\epsilon_0$ is the permittivity of vacuum. The Coulomb force couples the motion of the ions, allowing ion A to exchange momentum with ion B, which constantly dissipates kinetic energy via laser cooling. By this, kinetic energy can be indirectly removed from ion A, leading to a sympathetic cooling process.

For estimation of the damping coefficient $\beta_c$ we follow the semi-classical optical molasses theory of Doppler cooling~\cite{Phillips1989}: Under the assumption of a small velocity, ($\vert \Dot{\bm{x}}_{\text{B}}\vert\vert \bm{k}_{\text{L}}\vert\ll \gamma_0$, where $\vert \bm{k}_{\text{L}}\vert=2\pi/\lambda_{\text{L}}$, $\lambda_{\text{L}}$ is the cooling laser wavelength and $\gamma_0$ is the natural linewidth of the cooling transition) and moderate laser intensities $I_0$, the damping coefficient can be estimated by
\begin{equation}
    \beta_{\text{c}} =  4\hbar \vert \bm{k}_{\text{L}}\vert^2\dfrac{I_0}{I_{\text{sat}}}\dfrac{2\Delta/\gamma_0}{[1+6I_0/I_{\text{sat}}+(2\Delta/\gamma_0)^2]^2} \, ,
    \label{eq:damping_coefficient}
\end{equation}
in which $\Delta$ is the laser detuning from transition frequency $2\pi c_0/\lambda_{\text{L}}$ and $I_{\text{sat}}$ the saturation intensity dependent on the atomic transition~\cite{Phillips1989,Foot2004,Gould1997}. On resonance $\Delta=0$, $I_0/I_{\text{sat}}$ is the ratio of Rabi frequency $\omega_{\text{Rabi}}$ and decay rate $\gamma_0$ of spontaneous emission~\cite{Foot2004}: $I_0/I_{\text{sat}}=2\omega_{\text{Rabi}}^2/\gamma_0^2$.

In the denominator of Equation \ref{eq:damping_coefficient}, the saturation term $I_0/I_{\text{sat}}$ is multiplied by 6. This correction accounts for the two counter-propagating laser cooling beams of intensity $I_0$ along each of the three trapping axes~\cite{Phillips1989,Foot2004}.
Note also that this model assumes independence of the 6 cooling laser beams and neglects the Doppler limit.
\subsection{Evaluation of cooling performance}
\label{subsec:evaluation}
In order to evaluate the sympathetic cooling performance, the equations of motion of the two-ion system, Eqs. \ref{eq:eqm1} and \ref{eq:eqm2}, are solved numerically for $N_0$ initial conditions of ion A and B at each trap driving waveform $\Psi$ and different trap parameters $q$, while $a$ and $a_z$ are set constant. Each initial condition is simulated up to a defined maximum simulation time $t_0$, split into $n\in \mathbb{N}$ evaluation intervals of length $\Delta t$ such that $t_0=n\Delta t$.

After each evaluation interval, the time average $\langle\,\dots\rangle$ over $\Delta t$ of parameters such as ion distance $\langle R_{\text{AB}}\rangle =\langle\vert\bm{x}_{\text{A}}(t)-\bm{x}_{\text{B}}(t)\vert\rangle$ and temperature of coolant and sympathetic cooled ion $\langle T_{j}\rangle = \frac{2}{3k_{\text{B}}}\frac{m_j}{2}\langle \dot{\bm{x}}_{j}(t)^2\rangle$ ($j={\text{A}},{\text{B}}$) are computed. The micro-motion of the trapped particles is thereby averaged out since we focus on the mean cooling dynamic of the system. By using mean temperature and separation distance, we are able to determine the mean Coulomb-coupling parameter
\begin{equation}
    \Gamma= \dfrac{e^2}{4\pi\epsilon_0\langle R_{\text{AB}}\rangle k_{\text{B}}\langle T_{\text{A}}\rangle}
    \label{eq:gamma}
\end{equation}
of the two-ion system ($k_{\text{B}}$ is Boltzmann's constant). This dimensionless quantity, interpreted as the ratio of the Coulomb energy to thermal energy, allows for characterizing the state of a one-component plasma~\cite{Bonitz2008,Thompson2014,Drewsen2015}: When $\Gamma\ll 1$, the plasma is said to be weakly coupled and strongly coupled at $\Gamma>1$ respectively~\cite{Thompson2014}. At higher values at around $\Gamma>175$, a transition into a solid crystal structure, in case of trapped cooled ions known as a Coulomb-crystal, is observed~\cite{Pollock1973,Slattery1980,Farouki1993,Jones1996}.

This parameter is applied to the two-ion plasma as an indicator of its cooling state: The times $t_1,\,t_2,\,t_3$ at which $\Gamma^{(1)}\geq 0.001$, $\Gamma^{(2)}\geq 1$ and $\Gamma^{(3)}\geq 175$ was reached is saved for each initial condition. Let $N_{\Gamma}^{(k)}\leq N_0$ (with $k=1,2,3$) label the number of initial conditions that reached $\Gamma^{(k)}$ within simulation time $t_0$, then 'half life' cooling times $\tau_k$ are extracted numerically by defining $\int_0^{\tau_k}n^{(k)}(t)\,\text{d}t=N_{\Gamma}^{(k)}/2$, with $n^{(k)}(t)$ being the computed histogram distribution of $t_k$ of the sample. If $N_{\Gamma}^{(k)}=0$, $\tau_k$ is set to maximum simulation time $t_0$. Following this definition, e.g., $\tau_3$ indicates the time it took half of all detected crystallizations to crystallize. Furthermore, the cooling efficiency can be quantified using a pseudo cooling rate, $\eta^{(k)}=N_{\Gamma}^{(k)}/\tau_k$, which allows for benchmarking the cooling performance.

Four possible end-configurations were possible in the simulations:

(1) The formation of a \textit{Coulomb-crystal}, identified by a mean Coulomb-coupling parameter $\Gamma \geq 175$. If this $\Gamma$-value is passed, last $\langle T_{j}\rangle$ and $\langle R_{\text{AB}}\rangle$ are saved and the simulation stops. Below sufficient low temperatures ($\sim$ few mK), the two-ion system is found in a steady-state of lowest classical possible energy, in balance with the restoring ponderomotive trap force and Coulomb repulsion force~\cite{Bowe1999,Drewsen2015}. At Coulomb-crystallization, the equilibrium distance of a two-ion chain in a linear Paul trap aligned along the $z$-axis can be calculated by~\cite{Meyrath1998,James1998}
\begin{equation}
    R_{\text{eq}} = \Bigl(\dfrac{e z_0^2}{4\pi U_z\epsilon_0}\Bigr)^{1/3}.
    \label{eq:equilibrium_distance}
\end{equation}
Therefore, the system is expected to be in a final state with mean ion distance $\langle R_{\text{AB}}\rangle\approx R_{\text{eq}}$.

\begin{figure}[ht!]
\includegraphics[width=0.475\textwidth]{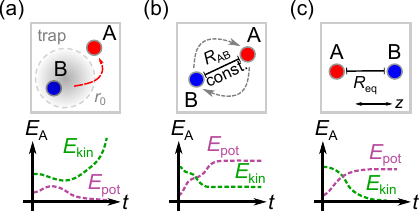}
\caption{The three possible end-states detected in the simulation and corresponding qualitative trend of mean potential and kinetic energy of ion A. The Coulomb parameter $\Gamma$ can be interpreted as their ratio, allowing for characterizing the states: (a) Unstable initial conditions, where ion A leaves the virtual trap volume if $\Gamma \to 0$ and/or $r_{\text{A}}>r_0$, (b) steady-state configuration of higher energies as in Coulomb-crystals, called orbit ($\Gamma \to const.$) and (c) crystallization of both ions along the axial trap axis, for $\Gamma \to \infty$.}
\label{fig:possible_states} 
\end{figure}

(2) \textit{Orbit states}. In our simulations, we observed steady-state ion trajectories of partially much higher kinetic energy compared to typical Coulomb-crystals. This means that the system remains in specific orbit configurations, manifesting themselves in a constant $\Gamma$ with constant temperatures $\langle T_{j}\rangle$ and ion distance $\langle R_{\text{AB}}\rangle$. Those frequency-locked orbit solutions have been reported in few papers as a phenomenon occurring in weakly damped two-ion systems described by non-linear coupled Mathieu-equations~\cite{Shen1997,Hoffnagle1993,Hoffnagle1994,Cardelli1994,Landa2012} similar to Eqs.~\ref{eq:eqm1}-\ref{eq:eqm2}. The dynamic undergoes transient chaos before trajectories merge into different attractors other than the Coulomb-crystal, dependent on trap voltage and damping magnitude $\beta_{\text{c}}$~\cite{Hoffnagle1994,Shen1997}. In experiments, those frequency locked-attractors are rarely observed, as the non-linear laser force quickly destabilizes such orbits~\cite{Cardelli1994,Hoffnagle1994}. To detect these states, the program tests for an approximately constant $\langle R_{\text{AB}}\rangle$ within a defined threshold time, magnitudes larger than one RF period: If the standard deviation of defined number of previously calculated values of $\langle R_{\text{AB}}\rangle$ combined is less than $0.5\,$µm, the simulation assumes a locked orbit and stops, while saving last $\langle R_{\text{AB}}\rangle$ and temperatures $\langle T_{j}\rangle$. The appropriate threshold time is estimated by manual testing as it depends on the initial conditions and maximum simulation time $t_0$.

(3) If $r_{\text{A}}(t)=\sqrt{x_{\text{A}}(t)^2+y_{\text{A}}(t)^2}>r_0$ or $z_{\text{A}}(t)>z_0$ within $\Delta t$, i.e., ion A left the simulated trap region, the simulation halts and the initial condition is classified as \textit{unstable}.

(4) Initial conditions, which ran up to the maximal simulation time $t_0$ without a triggered end-state, are classified as \textit{unidentified}.
\subsection{Simulation parameters}
\label{subsec:simulation_parameters}
In this subsection, trap parameters, initial conditions, and numerical parameters used to obtain the results in the next section are summarized.
\subsubsection{Choice of the two-ion system}
As an example of relevance, the sympathetic cooling ion A is chosen to be $^{27}$Al$^+$, since it is a promising candidate for optical atomic clocks and studied in various sympathetic cooling simulations as it lacks a technical easy accessible cooling transition~\cite{Chou2010,Wübenna2012,Hannig2019,Shang2016}. As particle B we selected the single ionized $^{26}$Mg$^+$ ion. The $^{26}$Mg$^+$ can be Doppler cooled using the well known 3$^2$S$_{1/2}$ $\to$ 3$^2$P$_{3/2}$ transition of wavelength $\lambda_{\text{L}} \approx 279.6\,$nm and can be easily produced~\cite{Batteiger2009,Hermann2009,Barrett2003,Schwarz2012}. 
The masses of the ions are $m_{\text{A}}=27\,$u, $m_{\text{B}}=24\,$u (with atomic mass unit $\text{u}=1.6605\times 10^{-27}\,$kg) and charges 1$e$ for Al-27 and Mg-26, respectively, which results in a mass ratio of $m_{\text{A}}/m_{\text{B}} \approx 1.038$.
Hence, the ions possess similar charge-to-mass ratios, which provides two advantages: Firstly, maximizing momentum transfer during Coulomb collisions reduces cooling and simulation time~\cite{Wübenna2012}. Secondly, since the stability of the quadrupole field in a linear Paul trap depends on the charge-to-mass ratio, similarity among ions allows for choosing suitable trap parameters more easily.
\subsubsection{Trap and cooling laser parameters}
The virtual radial trap size is defined as $r_0=1\,$mm, the axial trap length as $z_0=2\,$mm. The driving frequency is fixed at $\Omega = 2\pi\times 5\,$MHz, offset voltage $U_0$ set to zero ($a=0$) and segmentation voltage is set to $U_z=5\,$V in all simulations, which leads to $a_{z,\text{A}}\approx 0.018, a_{z,\text{B}}\approx 0.019$ for the Al-$27$ and the Mg-$26$ respectively. Based on the given axial field properties, Eq.~\ref{eq:equilibrium_distance} yields an expected crystal size of $R_{\text{eq}}\approx 10.48\,\text{µm}$. For a Coulomb-coupling parameter greater 175, this would require a mean temperature of $\langle T_{\text{A}}\rangle<10\,$mK simultaneously.

\begin{figure}[ht!]
\includegraphics[width=0.475\textwidth]{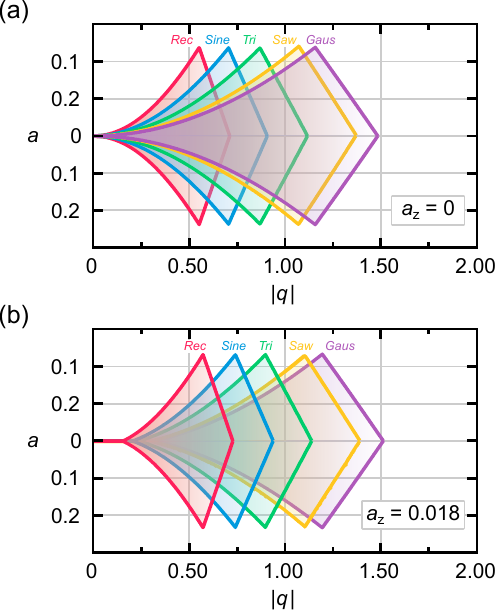}
\caption{Numerically calculated ideal stability diagram of a linear Paul trap and its dependence on the applied trap driving waveform. The colored areas represent trap parameters ($q$,$a$) at which radial confinement of an arbitrary particle was detected by the simulation. (a) Stability diagram without additional axial trapping field ($a_z = 0$) and (b) with $a_z = 0.018$ as given in the main simulations. A non-negligible axial field compresses and shifts the stability regions. See appendix Tab.~\ref{tab:table_stabilities} for detailed boundary values.}
\label{fig:stability_diagrams}
\end{figure}

The range of stable trapping parameters shifts depending on $\Psi$, and limits usable parameters in the main simulation. We therefore first computed ideal stability areas by numerically solving Eqs.~\ref{eq:hill_1} and \ref{eq:hill_2} for a single arbitrary particle with Runge-Kutta algorithm of 4th order for each studied RF waveform. The computed trap parameter range was set to $q\in[0,2]$, $a\in[-0.3,0.3]$ (step resolution $\Delta a=\Delta q = 0.001$, and 100 integration steps per RF period). A trap parameter tuple ($q,a$) is classified as stable if the particle's radial center distance $r(\xi)=\sqrt{x(\xi)^2+y(\xi)^2}$ stays confined (i.e., $r<r_0$) within simulation time of 100 RF periods. The particle initially was placed close to the trap center ($r(0)=r_0/1000$) with zero kinetic energy. The calculated ideal stability regions are mapped in Figure~\ref{fig:stability_diagrams}.

As can be observed, the stability area of an alternatively driven trap is shifted and squeezed compared to the sinusoidal-driven trap. By this numerical method, we obtain maximum stable $q$-parameter of $0.908$ in a sinusoidal, $0.712$ in a rectangular and $1.118$ in a triangular waveform driven trap, in agreement with reported boundary values~\cite{Ding2002,Ding2006,Brabeck2016,Xiong2012}. Furthermore, it can be observed that the maximum stable $q$-values of the non-traditional driving waveforms are scaled by their leading Fourier coefficients relative to the stability limit of the sinusoidal-driven trap, supporting previous observations~\cite{Brabeck2016B,Koizumi2009,Xiong2012,Bandelow2013,Reilly2015}. The maximum stable $q$-parameter at a square-wave driven trap can be obtained by multiplying the maximally stable $q$ at a sine-driven trap by $\pi/4$, likewise multiplying by $\pi^2/8$ yields maximum stable $q$ at the triangular-driven trap. This scaling also approximately holds true for the sawtooth wave with its leading Fourier coefficient $2/\pi$. Thereby, we assume a similar scaling manifests in the radial pseudo-potential depth $D_{r,\Psi}$ generated by each waveform (see also Refs.~\citenum{Reilly2015,Aksakal2016}). This scaling effect does not appear along the $a$-axis, as $a\neq0$ leads only to an equal offset voltage. It has to be mentioned that this geometrical condition of stability does not consider any influence due to the particles initial condition nor the influence of Coulomb interaction arising in the studied two-particle system. The defocusing components of the additional axial field lead to a shift of stable areas towards higher $q$-values and a slightly compressed stability area as can be viewed in Fig.~\ref{fig:stability_diagrams}(b). 

Due to this scaling of each stability region and hence different operating ranges along the $q$-axis depending on the trap driving waveform, we do not simulate an absolute $q$-parameter range, but set an individual $q$-parameter range for each waveform according to their stability boundaries in Fig.~\ref{fig:stability_diagrams}(b). Since $a=0=\textit{const.}$, only the $q$-value is varied in the simulations by changing the trap voltage $U_{\text{RF}}$. Both are related by $q=\alpha (U_{\text{RF}}/\text{V})$ with $\alpha = 2e/(m_{\text{A}}r_0^2\Omega^2)\cdot\text{V}\approx 0.00724$ at this given system parameters.

Considering the laser cooling transition line of $^{26}$Mg$^+$ at $\lambda_{\text{L}} = 279.6\,$nm~\cite{Batteiger2009,Hermann2009}, a detuning of $\Delta = -\gamma_0/2$ and a low intensity of $I_0/I_{\text{sat}}=0.1$, the damping coefficient $\beta_{\text{c}}$ is calculated to be $\beta_{\text{c}} \approx 3.151\times 10^{-21}\,\text{kgs}^{-1}$ or $\beta_{\text{c}}/m_{\text{B}} \approx 73\,011\,\text{s}^{-1}$ following Eq.~\ref{eq:damping_coefficient}. This magnitude of the damping coefficient is in line with the values commonly assumed in ion dynamic simulations~\cite{Schiller2003,Zhang2007}.
\subsubsection{Initial conditions and numerical parameters}
\begin{figure}[ht!]
\includegraphics[width=0.495\textwidth]{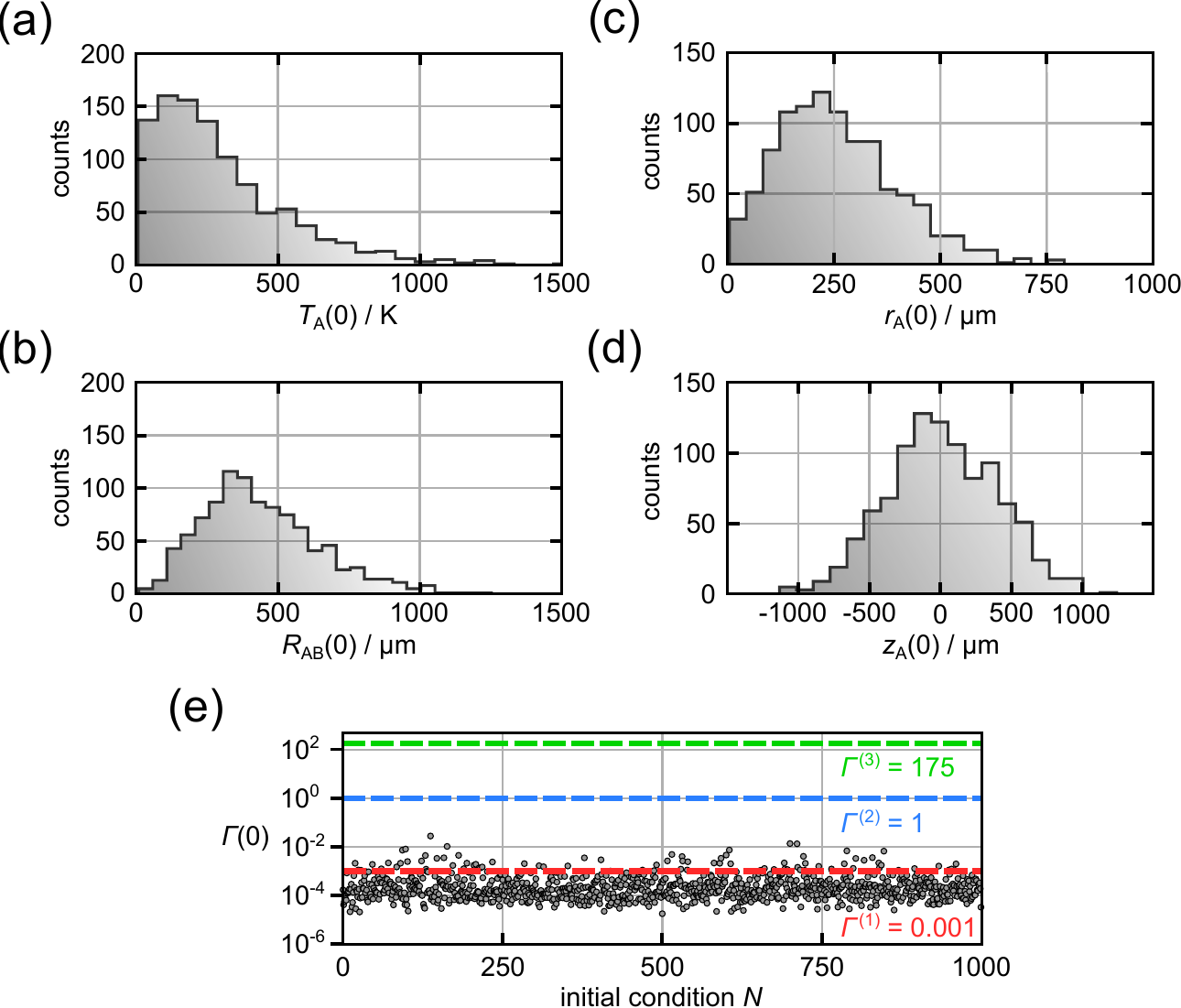}
\caption{\label{fig:initial_conditions} Histogram distributions of the initial condition sample of size $N_0=1000$ of sympathetic cooled $^{27}$Al$^+$ ions used in the simulations. Subplots show the (a) temperature, (b) separation distance between both ions, (c) radial center distance and (d) $z_{\text{A}}$-component at $t=0$. (e) Resulting distribution of initial Coulomb-coupling parameter $\Gamma$ and its boundaries of interest. With this given set of initial conditions, $\Gamma(0)$ lies below $10^{-3}$ for $97.4\%$ and below $10^{-4}$ for $42.0\%$ of $N_0$.}
\end{figure}

The sympathetic cooling duration is significantly dependent on the initial conditions of both particles. As a compromise between computation time and representative statistical result, a set of $N_0=1000$ random initial conditions was simulated. For the sake of simplicity, the coolant ion is initially placed in the trap center $x_{\text{B}}(0)=y_{\text{B}}(0)=z_{\text{B}}(0)=0$ with zero kinetic energy. Thereby, initial conditions refer to particle A only and their chosen distribution can be seen in Fig.~\ref{fig:initial_conditions}.
The initial position components of the $^{27}$Al$^+$ ions are randomly Gaussian distributed $\propto \exp{\bigl(-\frac{1}{2}(\frac{u}{\sigma_u})^2\bigr)}$ around zero with standard deviation $\sigma_u = r_0/5 =200\,\text{µm}$ ($u=x,y,z$), as shown exemplarily for the $z_{\text{A}}$ component in Fig.~\ref{fig:initial_conditions}(d). Each velocity component $v_u$ is randomized following the Maxwell-Boltzmann distribution~\cite{Maxwell1860},
\begin{equation}
        f_{\text{MB}}(v_u)= \left(\dfrac{m_\text{A}}{2 \pi k_{\text{B}}T}\right)^{1/2} \, \exp\left({ - \frac{m_\text{A} v_u^2}{2k_{\text{B}}T}}\right),
\end{equation}
at room-temperature ($T=293\,$K).

The sample of initial conditions is evaluated at 50 linear spaced $U_{\text{RF}}$-values along the stable $q$-axis of each driving waveform, following the boundaries from Tab.~\ref{tab:table_stabilities}.

Differential equations~\ref{eq:eqm1}-\ref{eq:eqm2} are numerically solved in Python by Runge-Kutta-algorithm of 4th order~\cite{Kutta1901}. Based on various testing on the ongoing cooling dynamics and computation time, 100 iteration steps during one RF period equivalent to a time step of $dt=2\,\text{ns}$ were chosen. This ensures not only a sufficient approximation of the solution of Eqs.~\ref{eq:eqm1}-\ref{eq:eqm2}, but also of the shape of studied driving waveforms. 

The total simulation time was set to $t_0=2000\,$ms and split into $n=20\,000$ evaluation intervals of a length of $\Delta t= 0.1\,$ms (500 RF periods). This simulation time is sufficient to keep track of the entire cooling process of the specified set of initial conditions. The minimum threshold time for detecting orbit states was set to $1/10\,t_0=200\,$ms. Therefore, the standard deviation of $2000$ values of $\langle R_{\text{AB}}\rangle$ is tested to be below $0.5\,$µm.
%%%%%%%%%%%%%%%%%%%%%%%%%%%%%%%%%%%%%%%%%%%%%
\section{Results and Discussion}
\begin{figure}[ht!]
\includegraphics[width=0.495\textwidth]{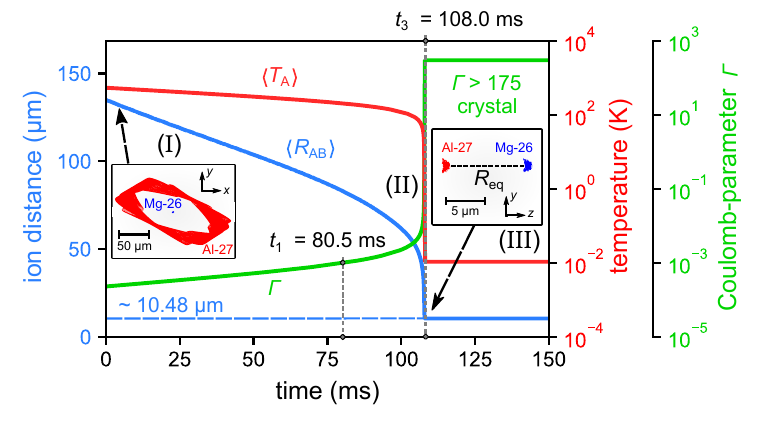}
\caption{\label{fig:ideal_example} Numerical solution of an initial condition leading to an ideal cooling process (trap driven by sinusoidal waveform). (I) At the beginning, there is only a low coupling of both ions resulting in a low transfer of momentum. RF-heating and trap force dominate the ions dynamics. The Al$^+$ ion's kinetic energy and distance to coolant ion Mg$^+$ decreases slowly, Coulomb parameter $\Gamma$ being in the range of  $\sim10^{-4}$ to $10^{-3}$. (II) After passing $t_1$, Coulomb interaction increases as the particles' distance decreases, thus more energy is extracted from $^{27}$Al$^+$; the cooling process accelerates and $\Gamma$ raises from $\sim10^{-2}$ to $>10^2$ within $1\,$ms. (III) Consequently, the end-state of a Coulomb-crystal with equilibrium ion distance of $10.48\,$µm is reached, due to the strong coupling. The simulation stops, saving $t_3 = 108\,$ms as crystallization time.}
\end{figure}

%%%%%%%%%%%%%%%%%%%%%%%%%%%%%%%%%%%%%%%%%%%%%
An ideal example of the behavior of the cooled $^{27}$Al$^+$  ion is shown in Figure~\ref{fig:ideal_example}.

The Al$^+$-Mg$^+$ system is initialized with an initial condition at a trap driving waveform $\Psi$. Generally, three phases can be distinguished: In the first phase, there is only a low coupling of both ions resulting in a low transfer of momentum, which in turn slowly reduces the average ion and trap center distance. The dynamic of the $^{27}$Al$^+$ ion is mainly dominated by the ponderomotive trapping force with low Coulomb perturbation. This phase of low energy transfer may spans up to several hundred milliseconds and $\Gamma$ typically lies below $10^{-3}$. Then, the Coulomb-collision rate increases as the particles' distance decreases, thus more energy is extracted from the sympathetically cooled particle and this process accelerates as the distance gets lower and lower. With a typical time span of less than $1$-$2\,$ms, this second phase marks the transition into the strong coupled regime, the last phase of the two-ion system being in a steady-state, ideally forming a Coulomb-crystal.
\subsection{Statistics of end-states}
\begin{figure}[ht!]
\includegraphics[width=0.475\textwidth]{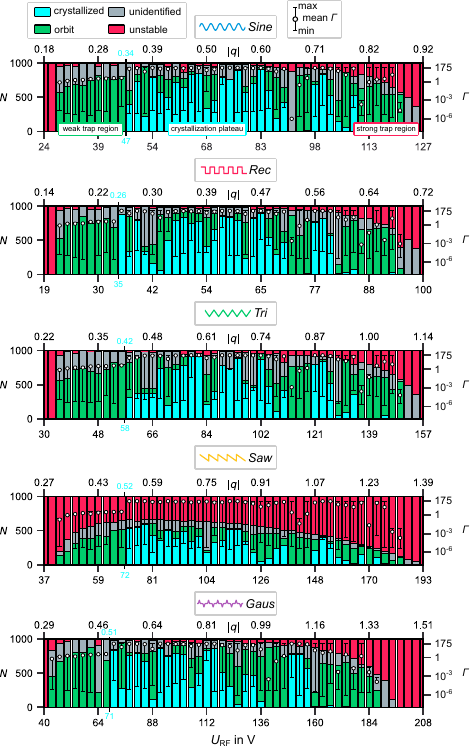}
\caption{\label{fig:endstate_statistics} Bar plot statistics of the resulting end-states of the Al$^+$-Mg$^+$ system in the simulations of each trap driving waveform $\Psi$ along their specific $q$-axis (50 simulated linear-spaced values). Each bar shows the fraction of appearances of the 4 possible end-states out of the set of initial conditions at a trap voltage $U_{\text{RF}}$ (lower x-axis) and its corresponding trap parameter $q$ (upper x-axis). The left y-axis labels the number of initial conditions up to $N_0=1000$. The right y-axis relates to the additional reduced box plots of each bar in which mean, maximum and minimum of Coulomb parameter $\Gamma$ are shown. Unstable and unidentified trajectories are excluded in the box plots data.} 
\end{figure}

Depending on the initial conditions, trap driving waveform and RF voltage the end-state of the ions varies. Figure~\ref{fig:endstate_statistics} shows the distribution between the four possible outcomes of the 1000 initial conditions with respect to the trap driving waveform $\Psi$ and along their individual stable $q$-axis. While varying $q$, we observed not only cooling dynamics into lowest energy state similar to Fig.~\ref{fig:ideal_example}, but also a significant number of transitions into higher energy states or even chaos due to random RF heating and system instabilities without any detected end-configuration within simulation time $t_0$.

\begin{figure}[ht!]
\includegraphics[width=.475\textwidth]{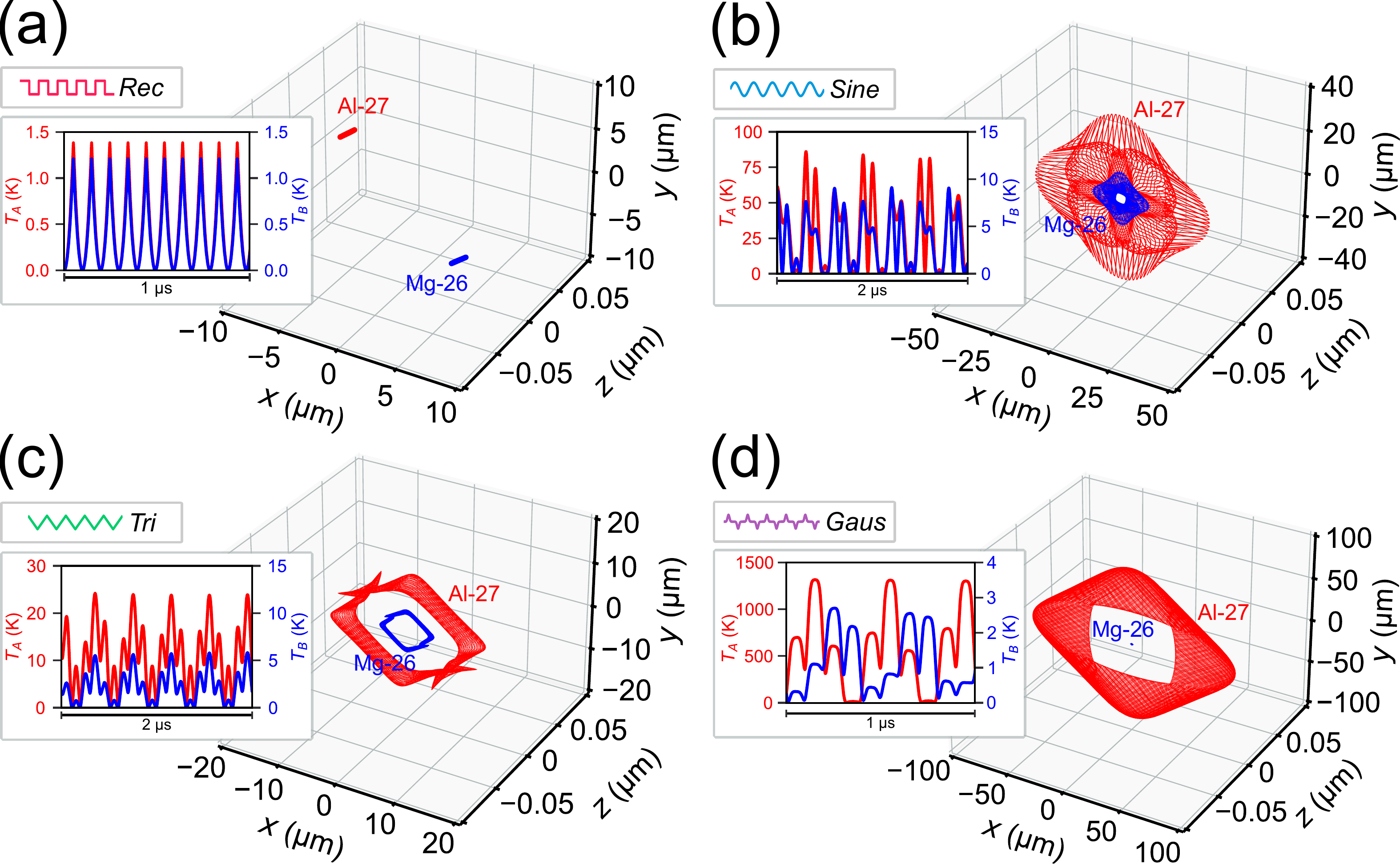}
\caption{\label{fig:orbits} Numerical solutions of orbit states of the Al$^+$-Mg$^+$ system with various kinetic- to potential energy range observed in the simulations. (a) A low energy state typically found at low trap voltages, where $\Gamma \approx 2.5$. (b-c) Frequently appeared orbit shapes with $\Gamma\approx 0.026$ and $\Gamma \approx 0.093$ at triangular and sinusoidal trap and moderate trapping force, (d) high energy steady-state with low Coulomb-coupling parameter of $\Gamma \approx 0.0005$.}
\end{figure}

Independent of the trap driving waveform, the resulting end-state distributions along the $q$-axis can be divided into 3 different regions: The \textit{weak trap region}, ranging from lowest stable trap voltage to the voltage, at which first Coulomb-crystallizations are detected. Although $N_{\Gamma}^{(3)} = 0$, this region characterizes a lot of orbit states of low total kinetic energy, usually $1 < \Gamma< 10$, for which a typical configuration is shown in Fig.~\ref{fig:orbits}(a). Both ions align in a crystal-like structure in the $x$-$y$-plane, driven by the ponderomotive trap force as the $z$-components are damped out. The equilibrium distance is in the range of $11$-$16\,$µm, but the kinetic energy of the $^{27}$Al$^+$ ion is not sufficiently low to form a Coulomb-crystal. Even though there is low energy gain of the particles due to the weak trapping force, the loose confinement leads to a low momentum exchange, therefore only negligible cooling of the $^{27}$Al$^+$ ion. Depending on the initial condition of the $^{27}$Al$^+$ ion, this might also be the reason for the high ratio of unidentified end-states in this weak potential region. In particular an $^{27}$Al$^+$ ion with initially high kinetic energy does not reach any end-state within the simulation time $t_0$. Raising the trap voltage $U_{\text{RF}}$ increases the rate of Coulomb collisions, promoting energy exchange until the lowest possible cooling state is reached; i.e., Coulomb crystals are observed in the statistics.

This threshold voltage depends on the applied trap driving waveform, but appears approximately in the same spot relative to each stability area and defines the beginning of the \textit{crystallization region}. It is marked in cyan on the axes in each subplot of Fig.~\ref{fig:endstate_statistics}. This middle region of moderate trapping force appears to be most suitable and efficient for sympathetic cooling as the statistics of end-states is dominated by crystallizations, together with low particle loss. We assume that the significant drop of crystallizations in the end-state statistics are correlated to more or less distinct system instabilities along the $q$-axis in the stability charts. The system gains energy, hence either reaching no end-state within $t_0$ or aligning in a high-energy orbit state with $\Gamma \ll 1$. Since the ideal model in Eq.~\ref{eq:trap_potential} does not involves any higher-order field perturbation, we assume they are not originated from well known non-linear resonances of the trap itself~\cite{Drakoudis2006,Alheit1995,Eades1992}, but rather from random two-ion collision instability heating~\cite{Harmon2003}. For instance, such an unwanted system instability is most clearly pronounced at $U_{\text{RF}}\approx 92\,$V ($q\approx 0.66$) at a sine-driven trap and appears nearly at the same position in each subplot of Fig.~\ref{fig:endstate_statistics}. The crystallization region should therefore be viewed as two plateaus, separated by an instability canyon. Based on the increase of the mean $\Gamma$-value within this canyon, it seems that again a threshold voltage has to be passed until lowest energy-state transitions are possible.

Finally, the third region spans roughly the last 1/5th of the stable $q$-axis, where the number of crystallizations reduces significantly to nearly zero ($N_{\Gamma}^{(3)}\ll 100$). By increasing the trap voltage up to the stability limit the trapping force dominates the sympathetic cooling process in most of the cases. The mean ion distance and RF heating increase, and a statistically significant number of trajectories are observed to be unstable until complete ion loss in all cases.

For the same reasons, orbit states are typically low coupled at the stability limit ($\Gamma \ll 1$) as exemplified in Fig.~\ref{fig:orbits}(d). It should be noted that the mean $\Gamma$ of orbit states increases within this strong trap region, indicating potentially again a transition phase into a third crystallization plateau, but halts upon reaching the stability limit. This can be especially observed in subplots of the sine, square and triangular waveform in Fig.~\ref{fig:endstate_statistics}.

\begin{figure}[ht!]
\includegraphics[width=0.475\textwidth]{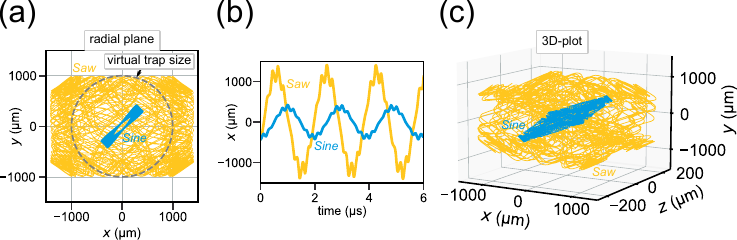}
\caption{\label{fig:unstable_sawtooth} Numerical trajectories of the $^{27}$Al$^+$ ion in first evaluation interval ($\Delta t=0.1\,$ms) with an identical initial condition, but at a sinusoidal- and sawtooth-driven linear Paul trap. (a) Movement in the $x$-$y$-plane. The gray circle indicates the virtual radial trap size $r_0$. (b) $x$-coordinate of $^{27}$Al$^+$ during initial 30 waveform periods. (c) 3D-view of the ion's motion. The trajectory of the $^{27}$Al$^+$ exceeds the virtual trap boundary in a sawtooth-driven trap, the initial condition is therefore detected to be unstable.}
\end{figure}

A remarkable difference in the end-state statistics is found in the number of unstable outcomes resulted at a sawtooth-driven Paul trap: While at a differently driven trap, the percentage of unstable trajectories can be minimized to less than 1$\,\%$ of $N_0$, a sawtooth-driven trap leads to a minimum of $\sim 33\,\%$ particle loss along the stable $q$-axis at this given set of initial conditions. The main reason lies in the generated particle dynamic and is illustrated by an example trajectory in Fig.~\ref{fig:unstable_sawtooth}. 

The sawtooth waveform generally produces substantially vast trajectories of the $^{27}$Al$^+$ ion compared to the other trap driving waveforms. Consequently, the instability condition $r>r_0$ is more frequently triggered as the $^{27}$Al$^+$ exits the trap volume, despite the motion being actually bounded.

Consequently, the quantity of unstable trajectories serves as an indirect measure of the confinement efficiency for each trap driving waveform, with the sawtooth waveform performing clearly the worst among the examined waveforms. Given the statistical results showing significant ion loss and fewer near-center passages, i.e., reduced Coulomb interaction between ions, supposes an inefficient sympathetic cooling in a sawtooth-driven trap.
\subsection{Sympathetic cooling efficiency}
\begin{figure}[ht!]
\includegraphics[width=0.495\textwidth]{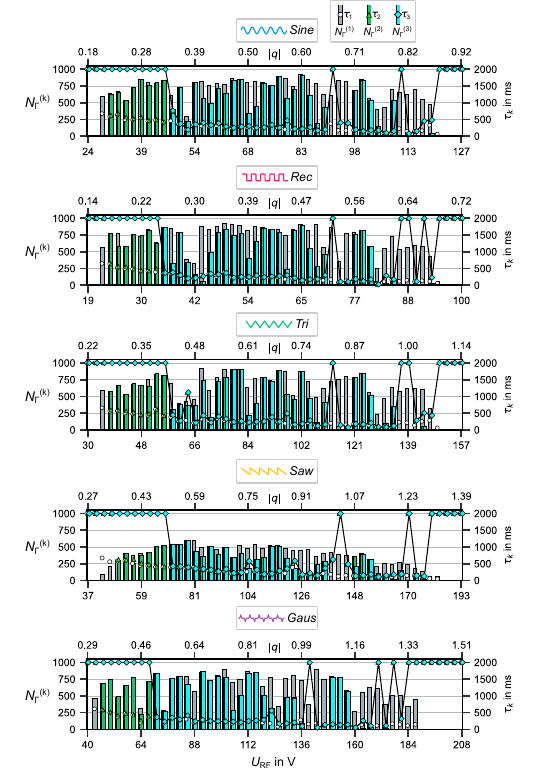}
\caption{\label{fig:tau_statistics} Numerical result of the sympathetic cooling times $\tau_k$ ($k=1,2,3$) of the $^{27}$Al$^+$ ion at each trap driving waveform $\Psi$ and their dependence on trap parameter $q$. Lower and upper x-axis label trap voltage and trap parameter respectively, with equal scaling as in Fig.~\ref{fig:endstate_statistics}. The bars show $N_{\Gamma}^{(k)}$, which labels the number of initial conditions that reached $\Gamma^{(k)}$ and symbols the corresponding cooling times $\tau_k$ whose values can be read from right-handed ordinate with $\Gamma^{(1)}\geq 0.001$, $\Gamma^{(2)}\geq 1$, and $\Gamma^{(3)}\geq 175$. Data points at $\tau_k = 2000~ \text{ms}$ mean that there were no instances the corresponding $\Gamma^{(k)}$ was reached.}
\end{figure}

In this section, we compare the sympathetic cooling efficiency of the studied trap driving waveforms by mainly focusing on the crystallization plateaus in Fig.~\ref{fig:endstate_statistics}. The cooling efficiency is defined by the ratio of cooling time $\tau_k$ and number of initial conditions that reached the chosen $\Gamma$-boundaries, $N_{\Gamma}^{(k)}$. Figure~\ref{fig:tau_statistics} shows both quantities separately for each examined trap driving waveform $\Psi$ along same trap voltage $U_{\text{RF}}$ and trap parameter $q$ range as in Fig.~\ref{fig:endstate_statistics}.

\begin{figure}[htbp]
\includegraphics[width=0.475\textwidth]{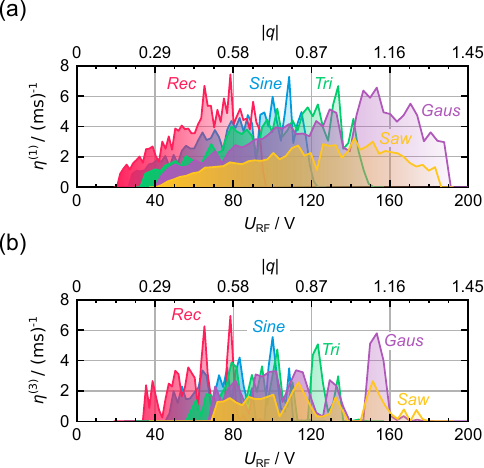}
\caption{\label{fig:cooling_rates} Sympathetic cooling efficiency $\eta$ of the $^{27}$Al$^+$ ion in dependency of trap driving waveform, trap parameter $q$ and trap voltage $U_{\text{RF}}$ as calculated from the simulations: (a) $\eta^{(1)}$, the efficiency of reaching the first boundary of Coulomb parameter ($\Gamma = 0.001$) and (b) $\eta^{(3)}$, of reaching the Coulomb-crystallization state ($\Gamma > 175$). A large value of $\eta$ indicates high cooling efficiency. Note that the distribution of $\eta^{(2)}$ differs insignificantly from that of $\eta^{(3)}$, and is therefore not shown.}
\end{figure}

The following general observations can be made: The cooling times $\tau_k$ decrease approximately linear with the trap voltage $U_{\text{RF}}$. When the trap potential becomes narrower, Coulomb-collisions are more probable, resulting in a faster extraction of kinetic energy from the $^{27}$Al$^+$ ion. On average, the cooling times are also slightly longer in a sawtooth-driven trap compared to the traditional driven trap, which is assumed to be due to the weak confinement. In the weak trap region, not only the longest cooling times $\tau_1$ and $\tau_2$ resulted, but a major part of the initial condition sample reaches $\Gamma >1$. There is also approximately less than $100\,$ms difference between $\tau_1$ and $\tau_2$, indicating a rapid cooling within the low coupled range $\Gamma<1$. The average cooling time $\tau_1$ in the weak trap region turns out to be highest using the sine waveform ($\sim 526\,$ms) and lowest using a Gaussian waveform ($\sim 416\,$ms). 

In contrast, operating in the strong trap region may lead to the lowest cooling times $\tau_1$ but the two-ion system stays at $\Gamma\ll1$ in most of the cases. Around half the sample (less than 1/4 at sawtooth waveform) quickly cools to $\Gamma^{(1)} = 10^{-3}$, but due to the dominant heating of the trap no tight ion coupling is achieved. Considering the initial cooling phase passing $\Gamma^{(1)}$, $N_{\Gamma}^{(1)}$ is not significantly lower compared to the statistics at lower trap voltage, but the cooling process appears to be faster: The average $\tau_1$ in this region is lowest with the sawtooth waveform ($\sim 103\,$ms) and highest with the sinusoidal waveform ($\sim 179\,$ms).

Despite the fact that the initial cooling is fastest near the stability limit, considering the clear increase in particle loss and high-energy orbits suggests operating further left on the $q$-axis to achieve effective sympathetic cooling of the $^{27}$Al$^+$ ion.

The statistically most efficient method for sympathetic cooling is to operate in the crystallization region, where the majority of $N_0$ reaches $\Gamma^{(3)} = 175$ and ion loss remains below 10\% (except when using a sawtooth wave, which results in a loss of over 33\%). It can be observed that cooling times $\tau_2$ and $\tau_3$ only differ by $\sim 1\,$ms. This is due to the fact that the transition from $\Gamma^{(2)} =1$ to $\Gamma^{(3)}=175$ runs fast compared to the first phase of cooling (see Fig.~\ref{fig:ideal_example}).

Excluding the instability canyons, the mean crystallization time $\tau_3$ in the first plateau is shortest with rectangular and Gaussian waveform with $\sim 266\,$ms and $\sim 267\,$ms respectively and longest with the triangular wave ($\sim 326\,$ms; for comparison, sine: $\sim 317\,$ms). Considering that the cooling times decrease with $q$, the results suggest operating at the end of the crystallization region, thus preferably in the second plateau. Despite the number of crystallizations $N_{\Gamma}^{(3)}$ is not significantly larger, the cooling times $\tau_3$ are shortest.

Figure~\ref{fig:cooling_rates} shows the cooling efficiency $\eta^{(k)}$, plotted in dependence of the trap waveform along the trap voltage range $0<U_{\text{RF}}<200\,$V. In Fig.~\ref{fig:cooling_rates}(a), the raising trend in cooling efficiency is mainly contributed by the linear decrease of $\tau_1$ with trap voltage $U_{\text{RF}}$. No significant differences of $\eta^{(1)}$ between the applied waveforms can be observed, except for the sawtooth wave. The low cooling efficiency of the sawtooth-driven trap is primarily compromised by the low number of stable trajectories and crystallizations, although having competitive cooling times $\tau_k$.

Sympathetic cooling of the $^{27}$Al$^+$ ion is most efficient at the second crystallization plateau, because a major part of the sample undergoes a direct smooth transition from $\Gamma(0)$ to $\Gamma^{(3)}=175$ at minimal cooling times $\tau_3$. Indeed, the spots of best sympathetic cooling efficiency are found mainly in this region (see Fig.~\ref{fig:cooling_rates}(c)) and the most important maxima of $\eta^{(3)}$ are given in Tab.~\ref{tab:table_besteta}. 

\begin{table}[ht]
\setlength\extrarowheight{2pt}
\caption{\label{tab:table_besteta}
Numerically calculated most efficient trap parameter $q$ and corresponding trap voltage $U_{\text{RF}}$ of each studied trap driving waveform for sympathetic cooling of the Al$^+$-Mg$^+$ system in the simulations. They are ranked from top to bottom based on the cooling efficiency factor $\eta^{(3)}$, defined as the number of detected Coulomb-crystallizations $N_{\Gamma}^{(3)}$ of a sample of 1000 initial conditions divided by their mean crystallization time $\tau_3$.}
\begin{ruledtabular}
\begin{tabular}{cccccc}
Waveform&\multicolumn{1}{c}{\textrm{$\eta^{(3)}$/(ms)$^{-1}$}}&
\multicolumn{1}{c}{\textrm{\textrm{$N_{\Gamma}^{(3)}$}}}&
\multicolumn{1}{c}{\textrm{$\tau_3$/ms}}&
\multicolumn{1}{c}{\textrm{$q$}}&
\multicolumn{1}{c}{\textrm{$U_{\text{RF}}$/V}}
\vspace{0.2cm}\\
\hline
 \textsl{Rec} & 6.93 &  819 & 118& 0.568 & 79 \\
 \textsl{Rec} & 6.26 & 902 & 144 & 0.473 & 65 \\
 \textsl{Gaus} & 5.79 & 777 & 134 & 1.110 & 153 \\
 \textsl{Sine} & 5.55 & 833 & 150 & 0.725 & 100 \\
 \textsl{Gaus} & 5.11 & 777 & 152 & 1.085 & 150\\
 \textsl{Tri} & 5.07 & 842 & 166 & 0.892 & 123 \\
 \textsl{Saw} & 2.64 & 397 & 150 & 1.100 & 151 \\
\end{tabular}
\end{ruledtabular}
\end{table}

Having a sinusoidal-driven Paul trap, the best cooling efficiency was reached at $q\approx 0.73$ with cooling time $\tau_3=150\,$ms and $\sim 83\%$ crystallization of the sample. In fact, three spots are found which outperform the traditional driving waveform: While having roughly the same crystallization yield ($\sim 82\%$), $\tau_3$ is $\sim 20\%$ lower in a rectangular-driven trap at trap parameter $q\approx0.57$. The cooling efficiency is also improved by $\sim 11\%$ at $q\approx 0.47$. A Gaussian-driven trap also appears to outperform the sine-driven trap, with $\eta^{(3)}$ maximizing at $q \approx 1.11$ in the second crystallization region, where $\sim 78\%$ of the sample and half of them crystallized after $\tau_3 = 134 \,$ms.

Although leading to a similar percentage of crystallization, sympathetic cooling in a triangular-driven trap tends to be less efficient because cooling times are $\sim 10$-$20\,$ms longer compared to the sinusoidal-driven trap.
A maximum of $\sim 50\%$ crystallization of the sample was able to be reached in a sawtooth-driven trap. The best cooling efficiency is also found at the second crystallization plateau at $q\approx 1.10$, with a competing cooling time of $\tau_3= 150\,$ms but merely $\sim 40\%$ crystallization of the sample.
%%%%%%%%%%%%%%%%%%%%%%%%%%%%%%%%%%%%%%%%%%%%%
\section{Conclusion}
%%%%%%%%%%%%%%%%%%%%%%%%%%%%%%%%%%%%%%%%%%%%%
With the goal to study possibly influences of the quadrupole trap driving waveform on the sympathetic cooling efficiency, a classical sympathetic cooling model, based on a two-ion system interacting via mutual Coulomb-force and laser force was numerically investigated. In this work, the cooling performance of the traditional sinusoidal-driven trap was compared to a rectangular-, triangular-, sawtooth- and Gaussian-driven trap. 

A statistical analysis over a range of initial conditions and trap parameters of the system showed a non-negligible influence of the trap driving waveform on the sympathetic cooling efficiency: At specific trap parameters, average cooling times of the sample could be reduced by $\sim 20\%$ in a rectangular-driven trap compared to the sinusoidal-driven Paul trap. A Gaussian-wave driven trap turned out to be also capable of outperforming the traditional waveform as it offered lower mean cooling times within a trap parameter range. While a triangular-driven Paul trap led to almost similar cooling performance compared to the sinusoidal-driven one, a sawtooth-driven trap showed significant ion loss along the stable trap parameter axis indicating the worst sympathetic cooling efficiency compared to all studied trap driving waveforms.

Furthermore, general observations of this classical two-ion sympathetic cooling system could be made: A specific threshold voltage has to be passed until transitions into lowest classical possible energy state, the Coulomb-crystal, are observed. Below this voltage, only low-energy orbit states were detected by the simulation. In addition, a significant drop of crystallizations in the statistics at certain trap voltages indicates system instabilities, which should be avoided. Generally, the sympathetic cooling times decrease approximately linear with increasing the trap voltage. However, since particle loss also increases with raising the trap voltage, the results suggest that operating at moderate trap parameters yields the best overall sympathetic cooling performance.

Numerical calculations of the ideal stability areas showed also a non-negligible dependence on the trap driving waveform and the well-known scaling effect of maximum stable trap parameters was consequently reproduced. Therefore, the presented simple numerical approach potentially allows for reliably calculating the ideal stability chart for any given trap driving waveform. Using a non-sinusoidal driven quadrupole trap enables adjustment of mass filtering, operable trap parameter range, and possibly improves mass resolution.

Still, open questions are implied by the results: How do trap waveforms perform in sympathetic cooling at different sets of initial conditions, laser cooling parameter, charge-to-mass ratio or more than two ions interacting in the trapping field? 

Also, from the different behavior of the cooling for different waveforms it can be speculated that the fastest and most efficient cooling could be achieved by a composite waveform that switches from one waveform to another in the different cooling stages. Further, the use of chirped waveforms might increase the sympathetic cooling efficiency.

The findings indicate potential benefits of further in-depth studies on the key properties that trap driving waveforms should possess to enhance the sympathetic cooling efficiency in Paul traps.
%%%%%%%%%%%%%%%%%%%%%%%%%%%%%%%%%%%%%%%%%%%%%
\newpage
%%%%%%%%%%%%%%%%%%%%%%%%%%%%%%%%%%%%%%%%%%%%%
\begin{acknowledgments}
This work was partially supported by the LUH compute cluster, which is funded by the Leibniz Universität Hannover, the Lower Saxony Ministry of Science and Culture (MWK) and the Deutsche Forschungsgemeinschaft (DFG, German Research Foundation). Data analysis was primarily carried out by Python packages NumPy~\cite{harris2020array} and SciPy~\cite{2020SciPy-NMeth}.
\end{acknowledgments}

\section*{Author declarations}
\subsection*{Conflict of interest}
The authors have no conflicts to disclose.
\subsection*{Author contributions}
OS carried out the simulations and wrote the code and manuscript draft. AWS supervised the study. All authors discussed the results and contributed to the interpretation. 
\section*{Data Availability}
The data that support the findings of this study are available from the corresponding author upon reasonable request.

\appendix

\section*{Appendix: Trap stability limits}
\begin{table}[h]
\setlength\extrarowheight{2pt}
\caption{\label{tab:table_stabilities}
Numerically calculated stability boundaries of trap parameters $q$ and $a$ of a linear Paul trap driven by different RF-waveforms with and without additional axial field influence. (a) Without axial field ($a_z = 0$), stability is given for $[-q_{\text{max}},q_{\text{max}}]$ (except trivial case, $q=0$). (b) For non-zero axial parameter, $\vert a_{\text{max}}\vert$ is similar to the values given in subtable (a), whereas $\vert q_{\text{min}}\vert$ changes.}
\begin{ruledtabular}
\begin{tabular}{ccccccc}
Waveform:&\multicolumn{1}{c}{\textsl{Sine}}&
\multicolumn{1}{c}{\textsl{Rec}}&
\multicolumn{1}{c}{\textsl{Tri}}&
\multicolumn{1}{c}{\textsl{Saw}}&
\multicolumn{1}{c}{\textsl{Gaus}} &
\vspace{0.1cm}\\
\hline
(a) $a_z=0\:\:\:\:\:\:\:\:$\\
$\vert q_{\text{max}}\vert$ &0.908& 0.712& 1.118& 1.372& 1.484\\
\vspace{1em}
$\vert a_{\text{max}}\vert\:$ &0.237& 0.237 & 0.236& 0.235 & 0.238 \\
\hline
(b) $a_z=0.018$\\
$\vert q_{\text{max}}\vert$ &0.923& 0.724 & 1.136& 1.395 & 1.509 \\
\vspace{1em}
$\vert q_{\text{min}}\vert\,$ &0.191& 0.149& 0.235& 0.290 & 0.313\\
\end{tabular}
\end{ruledtabular}
\end{table}

\bibliographystyle{apsrev4-2}
\bibliography{references}

\providecommand{\noopsort}[1]{}\providecommand{\singleletter}[1]{#1}%
\begin{thebibliography}{87}%
\makeatletter
\providecommand \@ifxundefined [1]{%
 \@ifx{#1\undefined}
}%
\providecommand \@ifnum [1]{%
 \ifnum #1\expandafter \@firstoftwo
 \else \expandafter \@secondoftwo
 \fi
}%
\providecommand \@ifx [1]{%
 \ifx #1\expandafter \@firstoftwo
 \else \expandafter \@secondoftwo
 \fi
}%
\providecommand \natexlab [1]{#1}%
\providecommand \enquote  [1]{``#1''}%
\providecommand \bibnamefont  [1]{#1}%
\providecommand \bibfnamefont [1]{#1}%
\providecommand \citenamefont [1]{#1}%
\providecommand \href@noop [0]{\@secondoftwo}%
\providecommand \href [0]{\begingroup \@sanitize@url \@href}%
\providecommand \@href[1]{\@@startlink{#1}\@@href}%
\providecommand \@@href[1]{\endgroup#1\@@endlink}%
\providecommand \@sanitize@url [0]{\catcode `\\12\catcode `\$12\catcode
  `\&12\catcode `\#12\catcode `\^12\catcode `\_12\catcode `\%12\relax}%
\providecommand \@@startlink[1]{}%
\providecommand \@@endlink[0]{}%
\providecommand \url  [0]{\begingroup\@sanitize@url \@url }%
\providecommand \@url [1]{\endgroup\@href {#1}{\urlprefix }}%
\providecommand \urlprefix  [0]{URL }%
\providecommand \Eprint [0]{\href }%
\providecommand \doibase [0]{https://doi.org/}%
\providecommand \selectlanguage [0]{\@gobble}%
\providecommand \bibinfo  [0]{\@secondoftwo}%
\providecommand \bibfield  [0]{\@secondoftwo}%
\providecommand \translation [1]{[#1]}%
\providecommand \BibitemOpen [0]{}%
\providecommand \bibitemStop [0]{}%
\providecommand \bibitemNoStop [0]{.\EOS\space}%
\providecommand \EOS [0]{\spacefactor3000\relax}%
\providecommand \BibitemShut  [1]{\csname bibitem#1\endcsname}%
\let\auto@bib@innerbib\@empty
%</preamble>
\bibitem [{\citenamefont {Dehmelt}(1990)}]{Dehmelt1990}%
  \BibitemOpen
  \bibfield  {author} {\bibinfo {author} {\bibfnamefont {H.}~\bibnamefont
  {Dehmelt}},\ }\href@noop {} {\bibfield  {journal} {\bibinfo  {journal}
  {Reviews of modern physics}\ }\textbf {\bibinfo {volume} {62}},\ \bibinfo
  {pages} {525} (\bibinfo {year} {1990})}\BibitemShut {NoStop}%
\bibitem [{\citenamefont {Cirac}\ and\ \citenamefont
  {Zoller}(1995)}]{Cirac1995}%
  \BibitemOpen
  \bibfield  {author} {\bibinfo {author} {\bibfnamefont {J.~I.}\ \bibnamefont
  {Cirac}}\ and\ \bibinfo {author} {\bibfnamefont {P.}~\bibnamefont {Zoller}},\
  }\href@noop {} {\bibfield  {journal} {\bibinfo  {journal} {Physical review
  letters}\ }\textbf {\bibinfo {volume} {74}},\ \bibinfo {pages} {4091}
  (\bibinfo {year} {1995})}\BibitemShut {NoStop}%
\bibitem [{\citenamefont {Schmidt-Kaler}\ \emph {et~al.}(2003)\citenamefont
  {Schmidt-Kaler}, \citenamefont {H{\"a}ffner}, \citenamefont {Riebe},
  \citenamefont {Gulde}, \citenamefont {Lancaster}, \citenamefont {Deuschle},
  \citenamefont {Becher}, \citenamefont {Roos}, \citenamefont {Eschner},\ and\
  \citenamefont {Blatt}}]{Schmidt-Kaler2003}%
  \BibitemOpen
  \bibfield  {author} {\bibinfo {author} {\bibfnamefont {F.}~\bibnamefont
  {Schmidt-Kaler}}, \bibinfo {author} {\bibfnamefont {H.}~\bibnamefont
  {H{\"a}ffner}}, \bibinfo {author} {\bibfnamefont {M.}~\bibnamefont {Riebe}},
  \bibinfo {author} {\bibfnamefont {S.}~\bibnamefont {Gulde}}, \bibinfo
  {author} {\bibfnamefont {G.~P.}\ \bibnamefont {Lancaster}}, \bibinfo {author}
  {\bibfnamefont {T.}~\bibnamefont {Deuschle}}, \bibinfo {author}
  {\bibfnamefont {C.}~\bibnamefont {Becher}}, \bibinfo {author} {\bibfnamefont
  {C.~F.}\ \bibnamefont {Roos}}, \bibinfo {author} {\bibfnamefont
  {J.}~\bibnamefont {Eschner}},\ and\ \bibinfo {author} {\bibfnamefont
  {R.}~\bibnamefont {Blatt}},\ }\href@noop {} {\bibfield  {journal} {\bibinfo
  {journal} {Nature}\ }\textbf {\bibinfo {volume} {422}},\ \bibinfo {pages}
  {408} (\bibinfo {year} {2003})}\BibitemShut {NoStop}%
\bibitem [{\citenamefont {Bruzewicz}\ \emph {et~al.}(2019)\citenamefont
  {Bruzewicz}, \citenamefont {Chiaverini}, \citenamefont {McConnell},\ and\
  \citenamefont {Sage}}]{Bruzewicz2019}%
  \BibitemOpen
  \bibfield  {author} {\bibinfo {author} {\bibfnamefont {C.~D.}\ \bibnamefont
  {Bruzewicz}}, \bibinfo {author} {\bibfnamefont {J.}~\bibnamefont
  {Chiaverini}}, \bibinfo {author} {\bibfnamefont {R.}~\bibnamefont
  {McConnell}},\ and\ \bibinfo {author} {\bibfnamefont {J.~M.}\ \bibnamefont
  {Sage}},\ }\href@noop {} {\bibfield  {journal} {\bibinfo  {journal} {Applied
  Physics Reviews}\ }\textbf {\bibinfo {volume} {6}},\ \bibinfo {pages}
  {021314} (\bibinfo {year} {2019})}\BibitemShut {NoStop}%
\bibitem [{\citenamefont {Friis}\ \emph {et~al.}(2018)\citenamefont {Friis},
  \citenamefont {Marty}, \citenamefont {Maier}, \citenamefont {Hempel},
  \citenamefont {Holz{\"{a}}pfel}, \citenamefont {Jurcevic}, \citenamefont
  {Plenio}, \citenamefont {Huber}, \citenamefont {Roos}, \citenamefont
  {Blatt},\ and\ \citenamefont {Lanyon}}]{Friis2018}%
  \BibitemOpen
  \bibfield  {author} {\bibinfo {author} {\bibfnamefont {N.}~\bibnamefont
  {Friis}}, \bibinfo {author} {\bibfnamefont {O.}~\bibnamefont {Marty}},
  \bibinfo {author} {\bibfnamefont {C.}~\bibnamefont {Maier}}, \bibinfo
  {author} {\bibfnamefont {C.}~\bibnamefont {Hempel}}, \bibinfo {author}
  {\bibfnamefont {M.}~\bibnamefont {Holz{\"{a}}pfel}}, \bibinfo {author}
  {\bibfnamefont {P.}~\bibnamefont {Jurcevic}}, \bibinfo {author}
  {\bibfnamefont {M.~B.}\ \bibnamefont {Plenio}}, \bibinfo {author}
  {\bibfnamefont {M.}~\bibnamefont {Huber}}, \bibinfo {author} {\bibfnamefont
  {C.}~\bibnamefont {Roos}}, \bibinfo {author} {\bibfnamefont {R.}~\bibnamefont
  {Blatt}},\ and\ \bibinfo {author} {\bibfnamefont {B.}~\bibnamefont
  {Lanyon}},\ }\href {https://doi.org/10.1103/PhysRevX.8.021012} {\bibfield
  {journal} {\bibinfo  {journal} {Physical Review X}\ }\textbf {\bibinfo
  {volume} {8}},\ \bibinfo {pages} {021012} (\bibinfo {year}
  {2018})}\BibitemShut {NoStop}%
\bibitem [{\citenamefont {Chou}\ \emph {et~al.}(2010)\citenamefont {Chou},
  \citenamefont {Hume}, \citenamefont {Koelemeij}, \citenamefont {Wineland},\
  and\ \citenamefont {Rosenband}}]{Chou2010}%
  \BibitemOpen
  \bibfield  {author} {\bibinfo {author} {\bibfnamefont {C.~W.}\ \bibnamefont
  {Chou}}, \bibinfo {author} {\bibfnamefont {D.~B.}\ \bibnamefont {Hume}},
  \bibinfo {author} {\bibfnamefont {J.~C.~J.}\ \bibnamefont {Koelemeij}},
  \bibinfo {author} {\bibfnamefont {D.~J.}\ \bibnamefont {Wineland}},\ and\
  \bibinfo {author} {\bibfnamefont {T.}~\bibnamefont {Rosenband}},\ }\href@noop
  {} {\bibfield  {journal} {\bibinfo  {journal} {Physical review letters}\
  }\textbf {\bibinfo {volume} {104}},\ \bibinfo {pages} {070802} (\bibinfo
  {year} {2010})}\BibitemShut {NoStop}%
\bibitem [{\citenamefont {Brewer}\ \emph {et~al.}(2019)\citenamefont {Brewer},
  \citenamefont {Chen}, \citenamefont {Hankin}, \citenamefont {Clements},
  \citenamefont {Chou}, \citenamefont {Wineland}, \citenamefont {Hume},\ and\
  \citenamefont {Leibrandt}}]{Brewer2019}%
  \BibitemOpen
  \bibfield  {author} {\bibinfo {author} {\bibfnamefont {S.~M.}\ \bibnamefont
  {Brewer}}, \bibinfo {author} {\bibfnamefont {J.-S.}\ \bibnamefont {Chen}},
  \bibinfo {author} {\bibfnamefont {A.~M.}\ \bibnamefont {Hankin}}, \bibinfo
  {author} {\bibfnamefont {E.~R.}\ \bibnamefont {Clements}}, \bibinfo {author}
  {\bibfnamefont {C.~W.}\ \bibnamefont {Chou}}, \bibinfo {author}
  {\bibfnamefont {D.~J.}\ \bibnamefont {Wineland}}, \bibinfo {author}
  {\bibfnamefont {D.~B.}\ \bibnamefont {Hume}},\ and\ \bibinfo {author}
  {\bibfnamefont {D.~R.}\ \bibnamefont {Leibrandt}},\ }\href@noop {} {\bibfield
   {journal} {\bibinfo  {journal} {Physical review letters}\ }\textbf {\bibinfo
  {volume} {123}},\ \bibinfo {pages} {033201} (\bibinfo {year}
  {2019})}\BibitemShut {NoStop}%
\bibitem [{\citenamefont {Paul}\ and\ \citenamefont
  {Steinwedel}(1953)}]{Paul1953}%
  \BibitemOpen
  \bibfield  {author} {\bibinfo {author} {\bibfnamefont {W.}~\bibnamefont
  {Paul}}\ and\ \bibinfo {author} {\bibfnamefont {H.}~\bibnamefont
  {Steinwedel}},\ }\href@noop {} {\bibfield  {journal} {\bibinfo  {journal}
  {Zeitschrift f{\"u}r Naturforschung A}\ }\textbf {\bibinfo {volume} {8}},\
  \bibinfo {pages} {448} (\bibinfo {year} {1953})}\BibitemShut {NoStop}%
\bibitem [{\citenamefont {Neuhauser}\ \emph {et~al.}(1980)\citenamefont
  {Neuhauser}, \citenamefont {Hohenstatt}, \citenamefont {Toschek},\ and\
  \citenamefont {Dehmelt}}]{Neuhauser1980}%
  \BibitemOpen
  \bibfield  {author} {\bibinfo {author} {\bibfnamefont {W.}~\bibnamefont
  {Neuhauser}}, \bibinfo {author} {\bibfnamefont {M.}~\bibnamefont
  {Hohenstatt}}, \bibinfo {author} {\bibfnamefont {P.}~\bibnamefont
  {Toschek}},\ and\ \bibinfo {author} {\bibfnamefont {H.}~\bibnamefont
  {Dehmelt}},\ }\href@noop {} {\bibfield  {journal} {\bibinfo  {journal}
  {Physical Review A}\ }\textbf {\bibinfo {volume} {22}},\ \bibinfo {pages}
  {1137} (\bibinfo {year} {1980})}\BibitemShut {NoStop}%
\bibitem [{\citenamefont {Roos}\ \emph {et~al.}(1999)\citenamefont {Roos},
  \citenamefont {Zeiger}, \citenamefont {Rohde}, \citenamefont {N{\"a}gerl},
  \citenamefont {Eschner}, \citenamefont {Leibfried}, \citenamefont
  {Schmidt-Kaler},\ and\ \citenamefont {Blatt}}]{Roos1999}%
  \BibitemOpen
  \bibfield  {author} {\bibinfo {author} {\bibfnamefont {C.}~\bibnamefont
  {Roos}}, \bibinfo {author} {\bibfnamefont {T.}~\bibnamefont {Zeiger}},
  \bibinfo {author} {\bibfnamefont {H.}~\bibnamefont {Rohde}}, \bibinfo
  {author} {\bibfnamefont {H.}~\bibnamefont {N{\"a}gerl}}, \bibinfo {author}
  {\bibfnamefont {J.}~\bibnamefont {Eschner}}, \bibinfo {author} {\bibfnamefont
  {D.}~\bibnamefont {Leibfried}}, \bibinfo {author} {\bibfnamefont
  {F.}~\bibnamefont {Schmidt-Kaler}},\ and\ \bibinfo {author} {\bibfnamefont
  {R.}~\bibnamefont {Blatt}},\ }\href@noop {} {\bibfield  {journal} {\bibinfo
  {journal} {Physical Review Letters}\ }\textbf {\bibinfo {volume} {83}},\
  \bibinfo {pages} {4713} (\bibinfo {year} {1999})}\BibitemShut {NoStop}%
\bibitem [{\citenamefont {Larson}\ \emph {et~al.}(1986)\citenamefont {Larson},
  \citenamefont {Bergquist}, \citenamefont {Bollinger}, \citenamefont {Itano},\
  and\ \citenamefont {Wineland}}]{Larson1986}%
  \BibitemOpen
  \bibfield  {author} {\bibinfo {author} {\bibfnamefont {D.}~\bibnamefont
  {Larson}}, \bibinfo {author} {\bibfnamefont {J.~C.}\ \bibnamefont
  {Bergquist}}, \bibinfo {author} {\bibfnamefont {J.~J.}\ \bibnamefont
  {Bollinger}}, \bibinfo {author} {\bibfnamefont {W.~M.}\ \bibnamefont
  {Itano}},\ and\ \bibinfo {author} {\bibfnamefont {D.~J.}\ \bibnamefont
  {Wineland}},\ }\href@noop {} {\bibfield  {journal} {\bibinfo  {journal}
  {Physical review letters}\ }\textbf {\bibinfo {volume} {57}},\ \bibinfo
  {pages} {70} (\bibinfo {year} {1986})}\BibitemShut {NoStop}%
\bibitem [{\citenamefont {Itano}\ \emph {et~al.}(1995)\citenamefont {Itano},
  \citenamefont {Bergquist}, \citenamefont {Bollinger},\ and\ \citenamefont
  {Wineland}}]{Wineland1995}%
  \BibitemOpen
  \bibfield  {author} {\bibinfo {author} {\bibfnamefont {W.~M.}\ \bibnamefont
  {Itano}}, \bibinfo {author} {\bibfnamefont {J.~C.}\ \bibnamefont
  {Bergquist}}, \bibinfo {author} {\bibfnamefont {J.~J.}\ \bibnamefont
  {Bollinger}},\ and\ \bibinfo {author} {\bibfnamefont {D.~J.}\ \bibnamefont
  {Wineland}},\ }\href {https://doi.org/10.1088/0031-8949/1995/T59/013}
  {\bibfield  {journal} {\bibinfo  {journal} {Physica Scripta}\ }\textbf
  {\bibinfo {volume} {1995}},\ \bibinfo {pages} {106} (\bibinfo {year}
  {1995})}\BibitemShut {NoStop}%
\bibitem [{\citenamefont {Arnold}\ \emph {et~al.}(1984)\citenamefont {Arnold},
  \citenamefont {Neuman},\ and\ \citenamefont {Pluchino}}]{Arnold1984}%
  \BibitemOpen
  \bibfield  {author} {\bibinfo {author} {\bibfnamefont {S.}~\bibnamefont
  {Arnold}}, \bibinfo {author} {\bibfnamefont {M.}~\bibnamefont {Neuman}},\
  and\ \bibinfo {author} {\bibfnamefont {A.~B.}\ \bibnamefont {Pluchino}},\
  }\href@noop {} {\bibfield  {journal} {\bibinfo  {journal} {Optics letters}\
  }\textbf {\bibinfo {volume} {9}},\ \bibinfo {pages} {4} (\bibinfo {year}
  {1984})}\BibitemShut {NoStop}%
\bibitem [{\citenamefont {Schell}\ \emph {et~al.}(2017)\citenamefont {Schell},
  \citenamefont {Kuhlicke}, \citenamefont {Kewes},\ and\ \citenamefont
  {Benson}}]{Schell2017}%
  \BibitemOpen
  \bibfield  {author} {\bibinfo {author} {\bibfnamefont {A.~W.}\ \bibnamefont
  {Schell}}, \bibinfo {author} {\bibfnamefont {A.}~\bibnamefont {Kuhlicke}},
  \bibinfo {author} {\bibfnamefont {G.}~\bibnamefont {Kewes}},\ and\ \bibinfo
  {author} {\bibfnamefont {O.}~\bibnamefont {Benson}},\ }\href@noop {}
  {\bibfield  {journal} {\bibinfo  {journal} {ACS Photonics}\ }\textbf
  {\bibinfo {volume} {4}},\ \bibinfo {pages} {2719} (\bibinfo {year}
  {2017})}\BibitemShut {NoStop}%
\bibitem [{\citenamefont {Berthelot}\ and\ \citenamefont
  {Bonod}(2019)}]{Berthelot2019}%
  \BibitemOpen
  \bibfield  {author} {\bibinfo {author} {\bibfnamefont {J.}~\bibnamefont
  {Berthelot}}\ and\ \bibinfo {author} {\bibfnamefont {N.}~\bibnamefont
  {Bonod}},\ }\href@noop {} {\bibfield  {journal} {\bibinfo  {journal} {Optics
  letters}\ }\textbf {\bibinfo {volume} {44}},\ \bibinfo {pages} {1476}
  (\bibinfo {year} {2019})}\BibitemShut {NoStop}%
\bibitem [{\citenamefont {Kuhlicke}\ \emph {et~al.}(2014)\citenamefont
  {Kuhlicke}, \citenamefont {Schell}, \citenamefont {Zoll},\ and\ \citenamefont
  {Benson}}]{Kuhlicke2014}%
  \BibitemOpen
  \bibfield  {author} {\bibinfo {author} {\bibfnamefont {A.}~\bibnamefont
  {Kuhlicke}}, \bibinfo {author} {\bibfnamefont {A.~W.}\ \bibnamefont
  {Schell}}, \bibinfo {author} {\bibfnamefont {J.}~\bibnamefont {Zoll}},\ and\
  \bibinfo {author} {\bibfnamefont {O.}~\bibnamefont {Benson}},\ }\href@noop {}
  {\bibfield  {journal} {\bibinfo  {journal} {Applied Physics Letters}\
  }\textbf {\bibinfo {volume} {105}},\ \bibinfo {pages} {073101} (\bibinfo
  {year} {2014})}\BibitemShut {NoStop}%
\bibitem [{\citenamefont {Delord}\ \emph {et~al.}(2017)\citenamefont {Delord},
  \citenamefont {Nicolas}, \citenamefont {Schwab},\ and\ \citenamefont
  {H{\'e}tet}}]{Delord2017}%
  \BibitemOpen
  \bibfield  {author} {\bibinfo {author} {\bibfnamefont {T.}~\bibnamefont
  {Delord}}, \bibinfo {author} {\bibfnamefont {L.}~\bibnamefont {Nicolas}},
  \bibinfo {author} {\bibfnamefont {L.}~\bibnamefont {Schwab}},\ and\ \bibinfo
  {author} {\bibfnamefont {G.}~\bibnamefont {H{\'e}tet}},\ }\href@noop {}
  {\bibfield  {journal} {\bibinfo  {journal} {New Journal of Physics}\ }\textbf
  {\bibinfo {volume} {19}},\ \bibinfo {pages} {033031} (\bibinfo {year}
  {2017})}\BibitemShut {NoStop}%
\bibitem [{\citenamefont {Bykov}\ \emph {et~al.}(2023)\citenamefont {Bykov},
  \citenamefont {Dania}, \citenamefont {Goschin},\ and\ \citenamefont
  {Northup}}]{Bykov2023}%
  \BibitemOpen
  \bibfield  {author} {\bibinfo {author} {\bibfnamefont {D.~S.}\ \bibnamefont
  {Bykov}}, \bibinfo {author} {\bibfnamefont {L.}~\bibnamefont {Dania}},
  \bibinfo {author} {\bibfnamefont {F.}~\bibnamefont {Goschin}},\ and\ \bibinfo
  {author} {\bibfnamefont {T.~E.}\ \bibnamefont {Northup}},\ }\href@noop {}
  {\bibfield  {journal} {\bibinfo  {journal} {Optica}\ }\textbf {\bibinfo
  {volume} {10}},\ \bibinfo {pages} {438} (\bibinfo {year} {2023})}\BibitemShut
  {NoStop}%
\bibitem [{\citenamefont {March}(1997)}]{March1997}%
  \BibitemOpen
  \bibfield  {author} {\bibinfo {author} {\bibfnamefont {R.~E.}\ \bibnamefont
  {March}},\ }\href
  {https://doi.org/10.1002/(SICI)1096-9888(199704)32:4<351::AID-JMS512>3.0.CO;2-Y}
  {\bibinfo {title} {{An introduction to quadrupole ion trap mass
  spectrometry}}} (\bibinfo {year} {1997})\BibitemShut {NoStop}%
\bibitem [{\citenamefont {Bushev}\ \emph {et~al.}(2006)\citenamefont {Bushev},
  \citenamefont {Rotter}, \citenamefont {Wilson}, \citenamefont {Dubin},
  \citenamefont {Becher}, \citenamefont {Eschner}, \citenamefont {Blatt},
  \citenamefont {Steixner}, \citenamefont {Rabl},\ and\ \citenamefont
  {Zoller}}]{Bushev2006}%
  \BibitemOpen
  \bibfield  {author} {\bibinfo {author} {\bibfnamefont {P.}~\bibnamefont
  {Bushev}}, \bibinfo {author} {\bibfnamefont {D.}~\bibnamefont {Rotter}},
  \bibinfo {author} {\bibfnamefont {A.}~\bibnamefont {Wilson}}, \bibinfo
  {author} {\bibfnamefont {F.}~\bibnamefont {Dubin}}, \bibinfo {author}
  {\bibfnamefont {C.}~\bibnamefont {Becher}}, \bibinfo {author} {\bibfnamefont
  {J.}~\bibnamefont {Eschner}}, \bibinfo {author} {\bibfnamefont
  {R.}~\bibnamefont {Blatt}}, \bibinfo {author} {\bibfnamefont
  {V.}~\bibnamefont {Steixner}}, \bibinfo {author} {\bibfnamefont
  {P.}~\bibnamefont {Rabl}},\ and\ \bibinfo {author} {\bibfnamefont
  {P.}~\bibnamefont {Zoller}},\ }\href@noop {} {\bibfield  {journal} {\bibinfo
  {journal} {Physical review letters}\ }\textbf {\bibinfo {volume} {96}},\
  \bibinfo {pages} {043003} (\bibinfo {year} {2006})}\BibitemShut {NoStop}%
\bibitem [{\citenamefont {Conangla}\ \emph {et~al.}(2018)\citenamefont
  {Conangla}, \citenamefont {Schell}, \citenamefont {Rica},\ and\ \citenamefont
  {Quidant}}]{Conangla2018}%
  \BibitemOpen
  \bibfield  {author} {\bibinfo {author} {\bibfnamefont {G.~P.}\ \bibnamefont
  {Conangla}}, \bibinfo {author} {\bibfnamefont {A.~W.}\ \bibnamefont
  {Schell}}, \bibinfo {author} {\bibfnamefont {R.~A.}\ \bibnamefont {Rica}},\
  and\ \bibinfo {author} {\bibfnamefont {R.}~\bibnamefont {Quidant}},\
  }\href@noop {} {\bibfield  {journal} {\bibinfo  {journal} {Nano letters}\
  }\textbf {\bibinfo {volume} {18}},\ \bibinfo {pages} {3956} (\bibinfo {year}
  {2018})}\BibitemShut {NoStop}%
\bibitem [{\citenamefont {Nagornykh}\ \emph {et~al.}(2015)\citenamefont
  {Nagornykh}, \citenamefont {Coppock},\ and\ \citenamefont
  {Kane}}]{Nagornykh2015}%
  \BibitemOpen
  \bibfield  {author} {\bibinfo {author} {\bibfnamefont {P.}~\bibnamefont
  {Nagornykh}}, \bibinfo {author} {\bibfnamefont {J.~E.}\ \bibnamefont
  {Coppock}},\ and\ \bibinfo {author} {\bibfnamefont {B.}~\bibnamefont
  {Kane}},\ }\href@noop {} {\bibfield  {journal} {\bibinfo  {journal} {Applied
  Physics Letters}\ }\textbf {\bibinfo {volume} {106}},\ \bibinfo {pages}
  {244102} (\bibinfo {year} {2015})}\BibitemShut {NoStop}%
\bibitem [{\citenamefont {Trypogeorgos}\ and\ \citenamefont
  {Foot}(2016)}]{Trypo2016}%
  \BibitemOpen
  \bibfield  {author} {\bibinfo {author} {\bibfnamefont {D.}~\bibnamefont
  {Trypogeorgos}}\ and\ \bibinfo {author} {\bibfnamefont {C.~J.}\ \bibnamefont
  {Foot}},\ }\href
  {https://doi.org/10.1103/PHYSREVA.94.023609/FIGURES/5/MEDIUM} {\bibfield
  {journal} {\bibinfo  {journal} {Physical Review A}\ }\textbf {\bibinfo
  {volume} {94}},\ \bibinfo {pages} {023609} (\bibinfo {year}
  {2016})}\BibitemShut {NoStop}%
\bibitem [{\citenamefont {Meng}\ and\ \citenamefont {Du}(2021)}]{Meng2021}%
  \BibitemOpen
  \bibfield  {author} {\bibinfo {author} {\bibfnamefont {Y.}~\bibnamefont
  {Meng}}\ and\ \bibinfo {author} {\bibfnamefont {L.}~\bibnamefont {Du}},\
  }\href {https://doi.org/10.1140/EPJD/S10053-020-00015-1} {\bibfield
  {journal} {\bibinfo  {journal} {The European Physical Journal D 2021 75:1}\
  }\textbf {\bibinfo {volume} {75}},\ \bibinfo {pages} {1} (\bibinfo {year}
  {2021})}\BibitemShut {NoStop}%
\bibitem [{\citenamefont {Ding}\ and\ \citenamefont
  {Kumashiro}(2006)}]{Ding2006}%
  \BibitemOpen
  \bibfield  {author} {\bibinfo {author} {\bibfnamefont {L.}~\bibnamefont
  {Ding}}\ and\ \bibinfo {author} {\bibfnamefont {S.}~\bibnamefont
  {Kumashiro}},\ }\href {https://doi.org/10.1002/rcm.2253} {\bibfield
  {journal} {\bibinfo  {journal} {Rapid Communications in Mass Spectrometry}\
  }\textbf {\bibinfo {volume} {20}},\ \bibinfo {pages} {3} (\bibinfo {year}
  {2006})}\BibitemShut {NoStop}%
\bibitem [{\citenamefont {Bandelow}\ \emph {et~al.}(2013)\citenamefont
  {Bandelow}, \citenamefont {Marx},\ and\ \citenamefont
  {Schweikhard}}]{Bandelow2013}%
  \BibitemOpen
  \bibfield  {author} {\bibinfo {author} {\bibfnamefont {S.}~\bibnamefont
  {Bandelow}}, \bibinfo {author} {\bibfnamefont {G.}~\bibnamefont {Marx}},\
  and\ \bibinfo {author} {\bibfnamefont {L.}~\bibnamefont {Schweikhard}},\
  }\href {https://doi.org/10.1016/j.ijms.2012.12.013} {\bibfield  {journal}
  {\bibinfo  {journal} {International Journal of Mass Spectrometry}\ }\textbf
  {\bibinfo {volume} {336}},\ \bibinfo {pages} {47} (\bibinfo {year}
  {2013})}\BibitemShut {NoStop}%
\bibitem [{\citenamefont {Ding}\ \emph {et~al.}(2002)\citenamefont {Ding},
  \citenamefont {Sudakov},\ and\ \citenamefont {Kumashiro}}]{Ding2002}%
  \BibitemOpen
  \bibfield  {author} {\bibinfo {author} {\bibfnamefont {L.}~\bibnamefont
  {Ding}}, \bibinfo {author} {\bibfnamefont {M.}~\bibnamefont {Sudakov}},\ and\
  \bibinfo {author} {\bibfnamefont {S.}~\bibnamefont {Kumashiro}},\ }\href
  {https://doi.org/10.1016/S1387-3806(02)00921-1} {\bibfield  {journal}
  {\bibinfo  {journal} {International Journal of Mass Spectrometry}\ }\textbf
  {\bibinfo {volume} {221}},\ \bibinfo {pages} {117} (\bibinfo {year}
  {2002})}\BibitemShut {NoStop}%
\bibitem [{\citenamefont {Koizumi}\ \emph {et~al.}(2009)\citenamefont
  {Koizumi}, \citenamefont {Whitten}, \citenamefont {Reilly},\ and\
  \citenamefont {Koizumi}}]{Koizumi2009}%
  \BibitemOpen
  \bibfield  {author} {\bibinfo {author} {\bibfnamefont {H.}~\bibnamefont
  {Koizumi}}, \bibinfo {author} {\bibfnamefont {W.~B.}\ \bibnamefont
  {Whitten}}, \bibinfo {author} {\bibfnamefont {P.~T.}\ \bibnamefont
  {Reilly}},\ and\ \bibinfo {author} {\bibfnamefont {E.}~\bibnamefont
  {Koizumi}},\ }\href {https://doi.org/10.1016/J.IJMS.2009.06.011} {\bibfield
  {journal} {\bibinfo  {journal} {International Journal of Mass Spectrometry}\
  }\textbf {\bibinfo {volume} {286}},\ \bibinfo {pages} {64} (\bibinfo {year}
  {2009})}\BibitemShut {NoStop}%
\bibitem [{\citenamefont {Brancia}\ \emph {et~al.}(2010)\citenamefont
  {Brancia}, \citenamefont {McCullough}, \citenamefont {Entwistle},
  \citenamefont {Grossmann},\ and\ \citenamefont {Ding}}]{Brancia2010}%
  \BibitemOpen
  \bibfield  {author} {\bibinfo {author} {\bibfnamefont {F.~L.}\ \bibnamefont
  {Brancia}}, \bibinfo {author} {\bibfnamefont {B.}~\bibnamefont {McCullough}},
  \bibinfo {author} {\bibfnamefont {A.}~\bibnamefont {Entwistle}}, \bibinfo
  {author} {\bibfnamefont {J.~G.}\ \bibnamefont {Grossmann}},\ and\ \bibinfo
  {author} {\bibfnamefont {L.}~\bibnamefont {Ding}},\ }\href
  {https://doi.org/10.1016/J.JASMS.2010.05.003} {\bibfield  {journal} {\bibinfo
   {journal} {Journal of the American Society for Mass Spectrometry}\ }\textbf
  {\bibinfo {volume} {21}},\ \bibinfo {pages} {1530} (\bibinfo {year}
  {2010})}\BibitemShut {NoStop}%
\bibitem [{\citenamefont {Brabeck}\ \emph {et~al.}(2016)\citenamefont
  {Brabeck}, \citenamefont {Koizumi}, \citenamefont {Koizumi},\ and\
  \citenamefont {Reilly}}]{Brabeck2016}%
  \BibitemOpen
  \bibfield  {author} {\bibinfo {author} {\bibfnamefont {G.~F.}\ \bibnamefont
  {Brabeck}}, \bibinfo {author} {\bibfnamefont {H.}~\bibnamefont {Koizumi}},
  \bibinfo {author} {\bibfnamefont {E.}~\bibnamefont {Koizumi}},\ and\ \bibinfo
  {author} {\bibfnamefont {P.~T.}\ \bibnamefont {Reilly}},\ }\href
  {https://doi.org/10.1016/j.ijms.2016.04.002} {\bibfield  {journal} {\bibinfo
  {journal} {International Journal of Mass Spectrometry}\ }\textbf {\bibinfo
  {volume} {404}},\ \bibinfo {pages} {8} (\bibinfo {year} {2016})}\BibitemShut
  {NoStop}%
\bibitem [{\citenamefont {Reilly}\ and\ \citenamefont
  {Brabeck}(2015)}]{Reilly2015}%
  \BibitemOpen
  \bibfield  {author} {\bibinfo {author} {\bibfnamefont {P.~T.}\ \bibnamefont
  {Reilly}}\ and\ \bibinfo {author} {\bibfnamefont {G.~F.}\ \bibnamefont
  {Brabeck}},\ }\href {https://doi.org/10.1016/J.IJMS.2015.09.013} {\bibfield
  {journal} {\bibinfo  {journal} {International Journal of Mass Spectrometry}\
  }\textbf {\bibinfo {volume} {392}},\ \bibinfo {pages} {86} (\bibinfo {year}
  {2015})}\BibitemShut {NoStop}%
\bibitem [{\citenamefont {Brabeck}\ and\ \citenamefont
  {Reilly}(2016)}]{Brabeck2016B}%
  \BibitemOpen
  \bibfield  {author} {\bibinfo {author} {\bibfnamefont {G.~F.}\ \bibnamefont
  {Brabeck}}\ and\ \bibinfo {author} {\bibfnamefont {P.~T.}\ \bibnamefont
  {Reilly}},\ }\href {https://doi.org/10.1007/S13361-016-1358-4/FIGURES/5}
  {\bibfield  {journal} {\bibinfo  {journal} {Journal of the American Society
  for Mass Spectrometry}\ }\textbf {\bibinfo {volume} {27}},\ \bibinfo {pages}
  {1122} (\bibinfo {year} {2016})}\BibitemShut {NoStop}%
\bibitem [{\citenamefont {Xiong}\ \emph {et~al.}(2012)\citenamefont {Xiong},
  \citenamefont {Xu}, \citenamefont {Zhou}, \citenamefont {Wang}, \citenamefont
  {Tang}, \citenamefont {Chen}, \citenamefont {Peng}, \citenamefont {Chang},\
  and\ \citenamefont {Nie}}]{Xiong2012}%
  \BibitemOpen
  \bibfield  {author} {\bibinfo {author} {\bibfnamefont {C.}~\bibnamefont
  {Xiong}}, \bibinfo {author} {\bibfnamefont {G.}~\bibnamefont {Xu}}, \bibinfo
  {author} {\bibfnamefont {X.}~\bibnamefont {Zhou}}, \bibinfo {author}
  {\bibfnamefont {J.}~\bibnamefont {Wang}}, \bibinfo {author} {\bibfnamefont
  {Y.}~\bibnamefont {Tang}}, \bibinfo {author} {\bibfnamefont {R.}~\bibnamefont
  {Chen}}, \bibinfo {author} {\bibfnamefont {W.~P.}\ \bibnamefont {Peng}},
  \bibinfo {author} {\bibfnamefont {H.~C.}\ \bibnamefont {Chang}},\ and\
  \bibinfo {author} {\bibfnamefont {Z.}~\bibnamefont {Nie}},\ }\href
  {https://doi.org/10.1039/c2an15756j} {\bibfield  {journal} {\bibinfo
  {journal} {Analyst}\ }\textbf {\bibinfo {volume} {137}},\ \bibinfo {pages}
  {1199} (\bibinfo {year} {2012})}\BibitemShut {NoStop}%
\bibitem [{\citenamefont {Opa{\v{c}}i{\'{c}}}\ \emph
  {et~al.}(2018)\citenamefont {Opa{\v{c}}i{\'{c}}}, \citenamefont {Hoffman},
  \citenamefont {Gotlib}, \citenamefont {Clowers},\ and\ \citenamefont
  {Reilly}}]{Opacic2018}%
  \BibitemOpen
  \bibfield  {author} {\bibinfo {author} {\bibfnamefont {B.}~\bibnamefont
  {Opa{\v{c}}i{\'{c}}}}, \bibinfo {author} {\bibfnamefont {N.~M.}\ \bibnamefont
  {Hoffman}}, \bibinfo {author} {\bibfnamefont {Z.~P.}\ \bibnamefont {Gotlib}},
  \bibinfo {author} {\bibfnamefont {B.~H.}\ \bibnamefont {Clowers}},\ and\
  \bibinfo {author} {\bibfnamefont {P.~T.}\ \bibnamefont {Reilly}},\ }\href
  {https://doi.org/10.1007/S13361-018-2012-0/ASSET/IMAGES/MEDIUM/JS8B05687_0006.PNG}
  {\bibfield  {journal} {\bibinfo  {journal} {Journal of the American Society
  for Mass Spectrometry}\ }\textbf {\bibinfo {volume} {29}},\ \bibinfo {pages}
  {2081} (\bibinfo {year} {2018})}\BibitemShut {NoStop}%
\bibitem [{\citenamefont {Aksakal}(2016)}]{Aksakal2016}%
  \BibitemOpen
  \bibfield  {author} {\bibinfo {author} {\bibfnamefont {H.}~\bibnamefont
  {Aksakal}},\ }\href {https://doi.org/10.1016/j.ijms.2015.11.003} {\bibfield
  {journal} {\bibinfo  {journal} {International Journal of Mass Spectrometry}\
  }\textbf {\bibinfo {volume} {394}},\ \bibinfo {pages} {22} (\bibinfo {year}
  {2016})}\BibitemShut {NoStop}%
\bibitem [{\citenamefont {Aksakal}\ and\ \citenamefont
  {Mercanli}(2020)}]{Aksakal2020}%
  \BibitemOpen
  \bibfield  {author} {\bibinfo {author} {\bibfnamefont {H.}~\bibnamefont
  {Aksakal}}\ and\ \bibinfo {author} {\bibfnamefont {A.~S.}\ \bibnamefont
  {Mercanli}},\ }\bibfield  {journal} {\bibinfo  {journal} {European Physical
  Journal Plus}\ }\textbf {\bibinfo {volume} {135}},\ \href
  {https://doi.org/10.1140/epjp/s13360-019-00082-3}
  {10.1140/epjp/s13360-019-00082-3} (\bibinfo {year} {2020})\BibitemShut
  {NoStop}%
\bibitem [{\citenamefont {Phillips}\ \emph {et~al.}(1989)\citenamefont
  {Phillips}, \citenamefont {Rolston}, \citenamefont {Watts}, \citenamefont
  {Tanner}, \citenamefont {Lett},\ and\ \citenamefont
  {Westbrook}}]{Phillips1989}%
  \BibitemOpen
  \bibfield  {author} {\bibinfo {author} {\bibfnamefont {W.~D.}\ \bibnamefont
  {Phillips}}, \bibinfo {author} {\bibfnamefont {S.~L.}\ \bibnamefont
  {Rolston}}, \bibinfo {author} {\bibfnamefont {R.~N.}\ \bibnamefont {Watts}},
  \bibinfo {author} {\bibfnamefont {C.~E.}\ \bibnamefont {Tanner}}, \bibinfo
  {author} {\bibfnamefont {P.~D.}\ \bibnamefont {Lett}},\ and\ \bibinfo
  {author} {\bibfnamefont {C.~I.}\ \bibnamefont {Westbrook}},\ }\href
  {https://doi.org/10.1364/JOSAB.6.002084} {\bibfield  {journal} {\bibinfo
  {journal} {JOSA B, Vol. 6, Issue 11, pp. 2084-2107}\ }\textbf {\bibinfo
  {volume} {6}},\ \bibinfo {pages} {2084} (\bibinfo {year} {1989})}\BibitemShut
  {NoStop}%
\bibitem [{\citenamefont {Drewsen}\ \emph {et~al.}(2003)\citenamefont
  {Drewsen}, \citenamefont {Jensen}, \citenamefont {Kjaergaard}, \citenamefont
  {Lindballe}, \citenamefont {Mortensen}, \citenamefont {M{\o}lhave},\ and\
  \citenamefont {Voigt}}]{Drewsen2003}%
  \BibitemOpen
  \bibfield  {author} {\bibinfo {author} {\bibfnamefont {M.}~\bibnamefont
  {Drewsen}}, \bibinfo {author} {\bibfnamefont {I.~S.}\ \bibnamefont {Jensen}},
  \bibinfo {author} {\bibfnamefont {N.}~\bibnamefont {Kjaergaard}}, \bibinfo
  {author} {\bibfnamefont {J.}~\bibnamefont {Lindballe}}, \bibinfo {author}
  {\bibfnamefont {A.}~\bibnamefont {Mortensen}}, \bibinfo {author}
  {\bibfnamefont {K.}~\bibnamefont {M{\o}lhave}},\ and\ \bibinfo {author}
  {\bibfnamefont {D.}~\bibnamefont {Voigt}},\ }\href
  {https://doi.org/10.1088/0953-4075/36/3/310} {\bibfield  {journal} {\bibinfo
  {journal} {Journal of Physics B: Atomic, Molecular and Optical Physics}\
  }\textbf {\bibinfo {volume} {36}},\ \bibinfo {pages} {525} (\bibinfo {year}
  {2003})}\BibitemShut {NoStop}%
\bibitem [{\citenamefont {Schiller}\ and\ \citenamefont
  {L{\"{a}}mmerzahl}(2003)}]{Schiller2003}%
  \BibitemOpen
  \bibfield  {author} {\bibinfo {author} {\bibfnamefont {S.}~\bibnamefont
  {Schiller}}\ and\ \bibinfo {author} {\bibfnamefont {C.}~\bibnamefont
  {L{\"{a}}mmerzahl}},\ }\href@noop {} {\bibfield  {journal} {\bibinfo
  {journal} {Physical Review A}\ }\textbf {\bibinfo {volume} {68}} (\bibinfo
  {year} {2003})}\BibitemShut {NoStop}%
\bibitem [{\citenamefont {W{\"{u}}bbena}\ \emph {et~al.}(2012)\citenamefont
  {W{\"{u}}bbena}, \citenamefont {Amairi}, \citenamefont {Mandel},\ and\
  \citenamefont {Schmidt}}]{Wübenna2012}%
  \BibitemOpen
  \bibfield  {author} {\bibinfo {author} {\bibfnamefont {J.~B.}\ \bibnamefont
  {W{\"{u}}bbena}}, \bibinfo {author} {\bibfnamefont {S.}~\bibnamefont
  {Amairi}}, \bibinfo {author} {\bibfnamefont {O.}~\bibnamefont {Mandel}},\
  and\ \bibinfo {author} {\bibfnamefont {P.~O.}\ \bibnamefont {Schmidt}},\
  }\href {https://doi.org/10.1103/PhysRevA.85.043412} {\bibfield  {journal}
  {\bibinfo  {journal} {Physical Review A}\ }\textbf {\bibinfo {volume} {85}},\
  \bibinfo {pages} {43412} (\bibinfo {year} {2012})}\BibitemShut {NoStop}%
\bibitem [{\citenamefont {Wuerker}\ \emph {et~al.}(1959)\citenamefont
  {Wuerker}, \citenamefont {Shelton},\ and\ \citenamefont
  {Langmuir}}]{Wuerker1959}%
  \BibitemOpen
  \bibfield  {author} {\bibinfo {author} {\bibfnamefont {R.~F.}\ \bibnamefont
  {Wuerker}}, \bibinfo {author} {\bibfnamefont {H.}~\bibnamefont {Shelton}},\
  and\ \bibinfo {author} {\bibfnamefont {R.~V.}\ \bibnamefont {Langmuir}},\
  }\href {https://doi.org/10.1063/1.1735165} {\bibfield  {journal} {\bibinfo
  {journal} {Journal of Applied Physics}\ }\textbf {\bibinfo {volume} {30}},\
  \bibinfo {pages} {342} (\bibinfo {year} {1959})}\BibitemShut {NoStop}%
\bibitem [{\citenamefont {Paul}(1990)}]{Paul1990}%
  \BibitemOpen
  \bibfield  {author} {\bibinfo {author} {\bibfnamefont {W.}~\bibnamefont
  {Paul}},\ }\href {https://doi.org/10.1103/RevModPhys.62.531} {\bibfield
  {journal} {\bibinfo  {journal} {Reviews of Modern Physics}\ }\textbf
  {\bibinfo {volume} {62}},\ \bibinfo {pages} {531} (\bibinfo {year}
  {1990})}\BibitemShut {NoStop}%
\bibitem [{\citenamefont {Drewsen}\ and\ \citenamefont
  {Br{\o}ner}(2000)}]{Drewsen2000}%
  \BibitemOpen
  \bibfield  {author} {\bibinfo {author} {\bibfnamefont {M.}~\bibnamefont
  {Drewsen}}\ and\ \bibinfo {author} {\bibfnamefont {A.}~\bibnamefont
  {Br{\o}ner}},\ }\href {https://doi.org/10.1103/PhysRevA.62.045401} {\bibfield
   {journal} {\bibinfo  {journal} {Physical Review A}\ }\textbf {\bibinfo
  {volume} {62}},\ \bibinfo {pages} {045401} (\bibinfo {year}
  {2000})}\BibitemShut {NoStop}%
\bibitem [{\citenamefont {Berkeland}\ \emph {et~al.}(1998)\citenamefont
  {Berkeland}, \citenamefont {Miller}, \citenamefont {Bergquist}, \citenamefont
  {Itano},\ and\ \citenamefont {Wineland}}]{Berkeland1998}%
  \BibitemOpen
  \bibfield  {author} {\bibinfo {author} {\bibfnamefont {D.~J.}\ \bibnamefont
  {Berkeland}}, \bibinfo {author} {\bibfnamefont {J.~D.}\ \bibnamefont
  {Miller}}, \bibinfo {author} {\bibfnamefont {J.~C.}\ \bibnamefont
  {Bergquist}}, \bibinfo {author} {\bibfnamefont {W.~M.}\ \bibnamefont
  {Itano}},\ and\ \bibinfo {author} {\bibfnamefont {D.~J.}\ \bibnamefont
  {Wineland}},\ }\href {https://doi.org/10.1063/1.367318} {\bibfield  {journal}
  {\bibinfo  {journal} {Journal of Applied Physics}\ }\textbf {\bibinfo
  {volume} {83}},\ \bibinfo {pages} {5025} (\bibinfo {year}
  {1998})}\BibitemShut {NoStop}%
\bibitem [{\citenamefont {Wineland}\ \emph {et~al.}(1999)\citenamefont
  {Wineland}, \citenamefont {Monroe}, \citenamefont {Itano}, \citenamefont
  {King}, \citenamefont {Leibfried}, \citenamefont {Meekhof}, \citenamefont
  {Myatt},\ and\ \citenamefont {Wood}}]{Wineland1999}%
  \BibitemOpen
  \bibfield  {author} {\bibinfo {author} {\bibfnamefont {D.}~\bibnamefont
  {Wineland}}, \bibinfo {author} {\bibfnamefont {C.}~\bibnamefont {Monroe}},
  \bibinfo {author} {\bibfnamefont {W.}~\bibnamefont {Itano}}, \bibinfo
  {author} {\bibfnamefont {B.}~\bibnamefont {King}}, \bibinfo {author}
  {\bibfnamefont {D.}~\bibnamefont {Leibfried}}, \bibinfo {author}
  {\bibfnamefont {D.}~\bibnamefont {Meekhof}}, \bibinfo {author} {\bibfnamefont
  {C.}~\bibnamefont {Myatt}},\ and\ \bibinfo {author} {\bibfnamefont
  {C.}~\bibnamefont {Wood}},\ }\href {https://doi.org/10.1002/3527603093.CH3}
  {\bibfield  {journal} {\bibinfo  {journal} {Progress of Physics}\ }\textbf
  {\bibinfo {volume} {46}},\ \bibinfo {pages} {363} (\bibinfo {year}
  {1999})}\BibitemShut {NoStop}%
\bibitem [{\citenamefont {Hill}(1900)}]{Hill1900}%
  \BibitemOpen
  \bibfield  {author} {\bibinfo {author} {\bibfnamefont {G.~W.}\ \bibnamefont
  {Hill}},\ }\href {https://doi.org/10.1007/BF02417081} {\bibfield  {journal}
  {\bibinfo  {journal} {Acta Mathematica}\ }\textbf {\bibinfo {volume} {8}},\
  \bibinfo {pages} {1} (\bibinfo {year} {1900})}\BibitemShut {NoStop}%
\bibitem [{\citenamefont {Pipes}(2004)}]{Hill2004}%
  \BibitemOpen
  \bibfield  {author} {\bibinfo {author} {\bibfnamefont {L.~A.}\ \bibnamefont
  {Pipes}},\ }\href {https://doi.org/10.1063/1.1721400} {\bibfield  {journal}
  {\bibinfo  {journal} {Journal of Applied Physics}\ }\textbf {\bibinfo
  {volume} {24}},\ \bibinfo {pages} {902} (\bibinfo {year} {2004})}\BibitemShut
  {NoStop}%
\bibitem [{\citenamefont {Mathieu}(1868)}]{Mathieu1868}%
  \BibitemOpen
  \bibfield  {author} {\bibinfo {author} {\bibfnamefont {E.}~\bibnamefont
  {Mathieu}},\ }\href@noop {} {\bibfield  {journal} {\bibinfo  {journal}
  {Journal de Math{\'{e}}matiques Pures et Appliqu{\'{e}}es}\ } (\bibinfo
  {year} {1868})}\BibitemShut {NoStop}%
\bibitem [{\citenamefont {Meixner}\ and\ \citenamefont
  {Sch{\"{a}}fke}(1954)}]{Meixner1954}%
  \BibitemOpen
  \bibfield  {author} {\bibinfo {author} {\bibfnamefont {J.}~\bibnamefont
  {Meixner}}\ and\ \bibinfo {author} {\bibfnamefont {F.~W.}\ \bibnamefont
  {Sch{\"{a}}fke}},\ }\href {https://doi.org/10.1007/978-3-662-00941-3} {\emph
  {\bibinfo {title} {Mathieusche Funktionen und Sph{\"{a}}roidfunktionen}}}\
  (\bibinfo  {publisher} {Springer Berlin Heidelberg},\ \bibinfo {year}
  {1954})\BibitemShut {NoStop}%
\bibitem [{\citenamefont {Abramowitz}\ \emph {et~al.}(1988)\citenamefont
  {Abramowitz}, \citenamefont {Stegun},\ and\ \citenamefont
  {Romer}}]{Abramowitz1988}%
  \BibitemOpen
  \bibfield  {author} {\bibinfo {author} {\bibfnamefont {M.}~\bibnamefont
  {Abramowitz}}, \bibinfo {author} {\bibfnamefont {I.~A.}\ \bibnamefont
  {Stegun}},\ and\ \bibinfo {author} {\bibfnamefont {R.~H.}\ \bibnamefont
  {Romer}},\ }\bibfield  {journal} {\bibinfo  {journal} {American Journal of
  Physics}\ }\href {https://doi.org/10.1119/1.15378} {10.1119/1.15378}
  (\bibinfo {year} {1988})\BibitemShut {NoStop}%
\bibitem [{\citenamefont {Kovacic}\ \emph {et~al.}(2018)\citenamefont
  {Kovacic}, \citenamefont {Rand},\ and\ \citenamefont {{Mohamed
  Sah}}}]{Kovacic2018}%
  \BibitemOpen
  \bibfield  {author} {\bibinfo {author} {\bibfnamefont {I.}~\bibnamefont
  {Kovacic}}, \bibinfo {author} {\bibfnamefont {R.}~\bibnamefont {Rand}},\ and\
  \bibinfo {author} {\bibfnamefont {S.}~\bibnamefont {{Mohamed Sah}}},\
  }\bibfield  {journal} {\bibinfo  {journal} {Applied Mechanics Reviews}\
  }\textbf {\bibinfo {volume} {70}},\ \href {https://doi.org/10.1115/1.4039144}
  {10.1115/1.4039144} (\bibinfo {year} {2018})\BibitemShut {NoStop}%
\bibitem [{\citenamefont {Dehmelt}(1968)}]{Dehmelt1968}%
  \BibitemOpen
  \bibfield  {author} {\bibinfo {author} {\bibfnamefont {H.~G.}\ \bibnamefont
  {Dehmelt}},\ }\href {https://doi.org/10.1016/S0065-2199(08)60170-0}
  {\bibfield  {journal} {\bibinfo  {journal} {Advances in Atomic, Molecular and
  Optical Physics}\ }\textbf {\bibinfo {volume} {3}},\ \bibinfo {pages} {53}
  (\bibinfo {year} {1968})}\BibitemShut {NoStop}%
\bibitem [{\citenamefont {Major}\ \emph {et~al.}(2005)\citenamefont {Major},
  \citenamefont {Gheorghe},\ and\ \citenamefont {Werth}}]{Major2005}%
  \BibitemOpen
  \bibfield  {author} {\bibinfo {author} {\bibfnamefont {F.~G.}\ \bibnamefont
  {Major}}, \bibinfo {author} {\bibfnamefont {V.~N.}\ \bibnamefont
  {Gheorghe}},\ and\ \bibinfo {author} {\bibfnamefont {G.}~\bibnamefont
  {Werth}},\ }\href {https://doi.org/10.1007/b137836} {\emph {\bibinfo {title}
  {Charged Particle Traps}}}\ (\bibinfo  {publisher} {Springer-Verlag},\
  \bibinfo {year} {2005})\BibitemShut {NoStop}%
\bibitem [{\citenamefont {Guan}\ and\ \citenamefont
  {Marshall}(1994)}]{Guan1994}%
  \BibitemOpen
  \bibfield  {author} {\bibinfo {author} {\bibfnamefont {S.}~\bibnamefont
  {Guan}}\ and\ \bibinfo {author} {\bibfnamefont {A.~G.}\ \bibnamefont
  {Marshall}},\ }\href {https://doi.org/10.1016/1044-0305(94)85038-0}
  {\bibfield  {journal} {\bibinfo  {journal} {Journal of the American Society
  for Mass Spectrometry}\ }\textbf {\bibinfo {volume} {5}},\ \bibinfo {pages}
  {64} (\bibinfo {year} {1994})}\BibitemShut {NoStop}%
\bibitem [{\citenamefont {Zhang}\ \emph {et~al.}(2007)\citenamefont {Zhang},
  \citenamefont {Offenberg}, \citenamefont {Roth}, \citenamefont {Wilson},\
  and\ \citenamefont {Schiller}}]{Zhang2007}%
  \BibitemOpen
  \bibfield  {author} {\bibinfo {author} {\bibfnamefont {C.~B.}\ \bibnamefont
  {Zhang}}, \bibinfo {author} {\bibfnamefont {D.}~\bibnamefont {Offenberg}},
  \bibinfo {author} {\bibfnamefont {B.}~\bibnamefont {Roth}}, \bibinfo {author}
  {\bibfnamefont {M.~A.}\ \bibnamefont {Wilson}},\ and\ \bibinfo {author}
  {\bibfnamefont {S.}~\bibnamefont {Schiller}},\ }\href
  {https://doi.org/10.1103/PHYSREVA.76.012719/FIGURES/22/MEDIUM} {\bibfield
  {journal} {\bibinfo  {journal} {Physical Review A - Atomic, Molecular, and
  Optical Physics}\ }\textbf {\bibinfo {volume} {76}},\ \bibinfo {pages}
  {012719} (\bibinfo {year} {2007})}\BibitemShut {NoStop}%
\bibitem [{\citenamefont {Guggemos}\ \emph {et~al.}(2015)\citenamefont
  {Guggemos}, \citenamefont {Heinrich}, \citenamefont {Herrera-Sancho},
  \citenamefont {Blatt},\ and\ \citenamefont {Roos}}]{Guggemos2015}%
  \BibitemOpen
  \bibfield  {author} {\bibinfo {author} {\bibfnamefont {M.}~\bibnamefont
  {Guggemos}}, \bibinfo {author} {\bibfnamefont {D.}~\bibnamefont {Heinrich}},
  \bibinfo {author} {\bibfnamefont {O.~A.}\ \bibnamefont {Herrera-Sancho}},
  \bibinfo {author} {\bibfnamefont {R.}~\bibnamefont {Blatt}},\ and\ \bibinfo
  {author} {\bibfnamefont {C.~F.}\ \bibnamefont {Roos}},\ }\href
  {https://doi.org/10.1088/1367-2630/17/10/103001} {\bibfield  {journal}
  {\bibinfo  {journal} {New Journal of Physics}\ }\textbf {\bibinfo {volume}
  {17}},\ \bibinfo {pages} {103001} (\bibinfo {year} {2015})}\BibitemShut
  {NoStop}%
\bibitem [{\citenamefont {Foot}(2004)}]{Foot2004}%
  \BibitemOpen
  \bibfield  {author} {\bibinfo {author} {\bibfnamefont {C.}~\bibnamefont
  {Foot}},\ }\href
  {https://global.oup.com/academic/product/atomic-physics-9780198506966} {\emph
  {\bibinfo {title} {{Atomic Physics - C.J. Foot - Oxford University Press}}}}\
  (\bibinfo  {publisher} {Oxford University Press},\ \bibinfo {year}
  {2004})\BibitemShut {NoStop}%
\bibitem [{\citenamefont {Gould}(1997)}]{Gould1997}%
  \BibitemOpen
  \bibfield  {author} {\bibinfo {author} {\bibfnamefont {P.}~\bibnamefont
  {Gould}},\ }\href {https://doi.org/10.1119/1.18740} {\bibfield  {journal}
  {\bibinfo  {journal} {American Journal of Physics}\ }\textbf {\bibinfo
  {volume} {65}},\ \bibinfo {pages} {1120} (\bibinfo {year}
  {1997})}\BibitemShut {NoStop}%
\bibitem [{\citenamefont {Bonitz}\ \emph {et~al.}(2008)\citenamefont {Bonitz},
  \citenamefont {Ludwig},\ and\ \citenamefont {Baumgartner}}]{Bonitz2008}%
  \BibitemOpen
  \bibfield  {author} {\bibinfo {author} {\bibfnamefont {M.}~\bibnamefont
  {Bonitz}}, \bibinfo {author} {\bibfnamefont {P.}~\bibnamefont {Ludwig}},\
  and\ \bibinfo {author} {\bibfnamefont {H.}~\bibnamefont {Baumgartner}},\
  }\href {https://doi.org/10.1063/1.2839297} {\bibfield  {journal} {\bibinfo
  {journal} {Phys. Plasmas}\ }\textbf {\bibinfo {volume} {15}},\ \bibinfo
  {pages} {55704} (\bibinfo {year} {2008})}\BibitemShut {NoStop}%
\bibitem [{\citenamefont {Thompson}(2014)}]{Thompson2014}%
  \BibitemOpen
  \bibfield  {author} {\bibinfo {author} {\bibfnamefont {R.~C.}\ \bibnamefont
  {Thompson}},\ }\href {https://doi.org/10.1080/00107514.2014.989715}
  {\bibfield  {journal} {\bibinfo  {journal} {Contemporary Physics}\ }\textbf
  {\bibinfo {volume} {56}},\ \bibinfo {pages} {63} (\bibinfo {year}
  {2014})}\BibitemShut {NoStop}%
\bibitem [{\citenamefont {Drewsen}(2015)}]{Drewsen2015}%
  \BibitemOpen
  \bibfield  {author} {\bibinfo {author} {\bibfnamefont {M.}~\bibnamefont
  {Drewsen}},\ }\href {https://doi.org/10.1016/j.physb.2014.11.050} {\bibfield
  {journal} {\bibinfo  {journal} {Physica B: Physics of Condensed Matter}\
  }\textbf {\bibinfo {volume} {460}},\ \bibinfo {pages} {105} (\bibinfo {year}
  {2015})}\BibitemShut {NoStop}%
\bibitem [{\citenamefont {Pollock}\ and\ \citenamefont
  {Hansen}(1973)}]{Pollock1973}%
  \BibitemOpen
  \bibfield  {author} {\bibinfo {author} {\bibfnamefont {E.~L.}\ \bibnamefont
  {Pollock}}\ and\ \bibinfo {author} {\bibfnamefont {J.~P.}\ \bibnamefont
  {Hansen}},\ }\href {https://doi.org/10.1103/PhysRevA.8.3110} {\bibfield
  {journal} {\bibinfo  {journal} {Physical Review A}\ }\textbf {\bibinfo
  {volume} {8}},\ \bibinfo {pages} {3110} (\bibinfo {year} {1973})}\BibitemShut
  {NoStop}%
\bibitem [{\citenamefont {Slattery}\ \emph {et~al.}(1980)\citenamefont
  {Slattery}, \citenamefont {Doolen},\ and\ \citenamefont
  {Dewitt}}]{Slattery1980}%
  \BibitemOpen
  \bibfield  {author} {\bibinfo {author} {\bibfnamefont {W.~L.}\ \bibnamefont
  {Slattery}}, \bibinfo {author} {\bibfnamefont {G.~D.}\ \bibnamefont
  {Doolen}},\ and\ \bibinfo {author} {\bibfnamefont {H.~E.}\ \bibnamefont
  {Dewitt}},\ }\href {https://doi.org/10.1103/PhysRevA.21.2087} {\bibfield
  {journal} {\bibinfo  {journal} {Physical Review A}\ }\textbf {\bibinfo
  {volume} {21}},\ \bibinfo {pages} {2087} (\bibinfo {year}
  {1980})}\BibitemShut {NoStop}%
\bibitem [{\citenamefont {Farouki}\ and\ \citenamefont
  {Hamaguchi}(1993)}]{Farouki1993}%
  \BibitemOpen
  \bibfield  {author} {\bibinfo {author} {\bibfnamefont {R.~T.}\ \bibnamefont
  {Farouki}}\ and\ \bibinfo {author} {\bibfnamefont {S.}~\bibnamefont
  {Hamaguchi}},\ }\href {https://doi.org/10.1103/PhysRevE.47.4330} {\bibfield
  {journal} {\bibinfo  {journal} {Physical Review E}\ }\textbf {\bibinfo
  {volume} {47}},\ \bibinfo {pages} {4330} (\bibinfo {year}
  {1993})}\BibitemShut {NoStop}%
\bibitem [{\citenamefont {Jones}\ and\ \citenamefont
  {Ceperley}(1996)}]{Jones1996}%
  \BibitemOpen
  \bibfield  {author} {\bibinfo {author} {\bibfnamefont {M.~D.}\ \bibnamefont
  {Jones}}\ and\ \bibinfo {author} {\bibfnamefont {D.~M.}\ \bibnamefont
  {Ceperley}},\ }\href {https://doi.org/10.1103/PhysRevLett.76.4572} {\bibfield
   {journal} {\bibinfo  {journal} {Physical Review Letters}\ }\textbf {\bibinfo
  {volume} {76}},\ \bibinfo {pages} {4572} (\bibinfo {year}
  {1996})}\BibitemShut {NoStop}%
\bibitem [{\citenamefont {Bowe}\ \emph {et~al.}(1999)\citenamefont {Bowe},
  \citenamefont {Hornek{\ae}r}, \citenamefont {Brodersen}, \citenamefont
  {Drewsen}, \citenamefont {Hangst},\ and\ \citenamefont
  {Schiffer}}]{Bowe1999}%
  \BibitemOpen
  \bibfield  {author} {\bibinfo {author} {\bibfnamefont {P.}~\bibnamefont
  {Bowe}}, \bibinfo {author} {\bibfnamefont {L.}~\bibnamefont {Hornek{\ae}r}},
  \bibinfo {author} {\bibfnamefont {C.}~\bibnamefont {Brodersen}}, \bibinfo
  {author} {\bibfnamefont {M.}~\bibnamefont {Drewsen}}, \bibinfo {author}
  {\bibfnamefont {J.~S.}\ \bibnamefont {Hangst}},\ and\ \bibinfo {author}
  {\bibfnamefont {J.~P.}\ \bibnamefont {Schiffer}},\ }\href
  {https://doi.org/10.1103/PhysRevLett.82.2071} {\bibfield  {journal} {\bibinfo
   {journal} {Physical Review Letters}\ }\textbf {\bibinfo {volume} {82}},\
  \bibinfo {pages} {2071} (\bibinfo {year} {1999})}\BibitemShut {NoStop}%
\bibitem [{\citenamefont {Meyrath}\ and\ \citenamefont {{James
  '}}(1998)}]{Meyrath1998}%
  \BibitemOpen
  \bibfield  {author} {\bibinfo {author} {\bibfnamefont {T.~P.}\ \bibnamefont
  {Meyrath}}\ and\ \bibinfo {author} {\bibfnamefont {D.~F.~V.}\ \bibnamefont
  {{James '}}},\ }\href@noop {} {\bibfield  {journal} {\bibinfo  {journal}
  {Physics Letters A}\ }\textbf {\bibinfo {volume} {240}},\ \bibinfo {pages}
  {37} (\bibinfo {year} {1998})}\BibitemShut {NoStop}%
\bibitem [{\citenamefont {James}(1998)}]{James1998}%
  \BibitemOpen
  \bibfield  {author} {\bibinfo {author} {\bibfnamefont {D.~F.~V.}\
  \bibnamefont {James}},\ }\href@noop {} {\bibfield  {journal} {\bibinfo
  {journal} {Appl. Phys. B}\ }\textbf {\bibinfo {volume} {66}},\ \bibinfo
  {pages} {181} (\bibinfo {year} {1998})}\BibitemShut {NoStop}%
\bibitem [{\citenamefont {Shen}\ \emph {et~al.}(1997)\citenamefont {Shen},
  \citenamefont {Yin}, \citenamefont {Dai},\ and\ \citenamefont
  {Zhang}}]{Shen1997}%
  \BibitemOpen
  \bibfield  {author} {\bibinfo {author} {\bibfnamefont {J.~L.}\ \bibnamefont
  {Shen}}, \bibinfo {author} {\bibfnamefont {H.~W.}\ \bibnamefont {Yin}},
  \bibinfo {author} {\bibfnamefont {J.~H.}\ \bibnamefont {Dai}},\ and\ \bibinfo
  {author} {\bibfnamefont {H.~J.}\ \bibnamefont {Zhang}},\ }\href
  {https://doi.org/10.1103/PhysRevA.55.2159} {\bibfield  {journal} {\bibinfo
  {journal} {Physical Review A}\ }\textbf {\bibinfo {volume} {55}},\ \bibinfo
  {pages} {2159} (\bibinfo {year} {1997})}\BibitemShut {NoStop}%
\bibitem [{\citenamefont {Hoffnagle}\ and\ \citenamefont
  {Brewer}(1993)}]{Hoffnagle1993}%
  \BibitemOpen
  \bibfield  {author} {\bibinfo {author} {\bibfnamefont {J.}~\bibnamefont
  {Hoffnagle}}\ and\ \bibinfo {author} {\bibfnamefont {R.~G.}\ \bibnamefont
  {Brewer}},\ }\href {https://doi.org/10.1103/PhysRevLett.71.1828} {\bibfield
  {journal} {\bibinfo  {journal} {Physical Review Letters}\ }\textbf {\bibinfo
  {volume} {71}},\ \bibinfo {pages} {1828} (\bibinfo {year}
  {1993})}\BibitemShut {NoStop}%
\bibitem [{\citenamefont {Hoffnagle}\ and\ \citenamefont
  {Brewer}(1994)}]{Hoffnagle1994}%
  \BibitemOpen
  \bibfield  {author} {\bibinfo {author} {\bibfnamefont {J.}~\bibnamefont
  {Hoffnagle}}\ and\ \bibinfo {author} {\bibfnamefont {R.~G.}\ \bibnamefont
  {Brewer}},\ }\href {https://doi.org/10.1103/PhysRevA.50.4157} {\bibfield
  {journal} {\bibinfo  {journal} {Physical Review A}\ }\textbf {\bibinfo
  {volume} {50}},\ \bibinfo {pages} {4157} (\bibinfo {year}
  {1994})}\BibitemShut {NoStop}%
\bibitem [{\citenamefont {Cardelli}\ \emph {et~al.}(1994)\citenamefont
  {Cardelli}, \citenamefont {Ebbets}, \citenamefont {{Attaining High}},
  \citenamefont {Hoffnagle},\ and\ \citenamefont {Brewer}}]{Cardelli1994}%
  \BibitemOpen
  \bibfield  {author} {\bibinfo {author} {\bibfnamefont {J.~A.}\ \bibnamefont
  {Cardelli}}, \bibinfo {author} {\bibfnamefont {D.~C.}\ \bibnamefont
  {Ebbets}}, \bibinfo {author} {\bibfnamefont {I.}~\bibnamefont {{Attaining
  High}}}, \bibinfo {author} {\bibfnamefont {J.}~\bibnamefont {Hoffnagle}},\
  and\ \bibinfo {author} {\bibfnamefont {R.~G.}\ \bibnamefont {Brewer}},\
  }\href {https://doi.org/10.1126/SCIENCE.265.5169.213} {\bibfield  {journal}
  {\bibinfo  {journal} {Science}\ }\textbf {\bibinfo {volume} {265}},\ \bibinfo
  {pages} {213} (\bibinfo {year} {1994})}\BibitemShut {NoStop}%
\bibitem [{\citenamefont {Landa}\ \emph {et~al.}(2012)\citenamefont {Landa},
  \citenamefont {Drewsen}, \citenamefont {Reznik},\ and\ \citenamefont
  {Retzker}}]{Landa2012}%
  \BibitemOpen
  \bibfield  {author} {\bibinfo {author} {\bibfnamefont {H.}~\bibnamefont
  {Landa}}, \bibinfo {author} {\bibfnamefont {M.}~\bibnamefont {Drewsen}},
  \bibinfo {author} {\bibfnamefont {B.}~\bibnamefont {Reznik}},\ and\ \bibinfo
  {author} {\bibfnamefont {A.}~\bibnamefont {Retzker}},\ }\href
  {https://doi.org/10.1088/1367-2630/14/9/093023} {\bibfield  {journal}
  {\bibinfo  {journal} {New Journal of Physics}\ }\textbf {\bibinfo {volume}
  {14}},\ \bibinfo {pages} {093023} (\bibinfo {year} {2012})}\BibitemShut
  {NoStop}%
\bibitem [{\citenamefont {Hannig}\ \emph {et~al.}(2019)\citenamefont {Hannig},
  \citenamefont {Pelzer}, \citenamefont {Scharnhorst}, \citenamefont {Kramer},
  \citenamefont {Stepanova}, \citenamefont {Xu}, \citenamefont {Spethmann},
  \citenamefont {Leroux}, \citenamefont {Mehlst{\"{a}}ubler},\ and\
  \citenamefont {Schmidt}}]{Hannig2019}%
  \BibitemOpen
  \bibfield  {author} {\bibinfo {author} {\bibfnamefont {S.}~\bibnamefont
  {Hannig}}, \bibinfo {author} {\bibfnamefont {L.}~\bibnamefont {Pelzer}},
  \bibinfo {author} {\bibfnamefont {N.}~\bibnamefont {Scharnhorst}}, \bibinfo
  {author} {\bibfnamefont {J.}~\bibnamefont {Kramer}}, \bibinfo {author}
  {\bibfnamefont {M.}~\bibnamefont {Stepanova}}, \bibinfo {author}
  {\bibfnamefont {Z.~T.}\ \bibnamefont {Xu}}, \bibinfo {author} {\bibfnamefont
  {N.}~\bibnamefont {Spethmann}}, \bibinfo {author} {\bibfnamefont {I.~D.}\
  \bibnamefont {Leroux}}, \bibinfo {author} {\bibfnamefont {T.~E.}\
  \bibnamefont {Mehlst{\"{a}}ubler}},\ and\ \bibinfo {author} {\bibfnamefont
  {P.~O.}\ \bibnamefont {Schmidt}},\ }\href {https://doi.org/10.1063/1.5090583}
  {\bibfield  {journal} {\bibinfo  {journal} {Review of Scientific
  Instruments}\ }\textbf {\bibinfo {volume} {90}},\ \bibinfo {pages} {53204}
  (\bibinfo {year} {2019})}\BibitemShut {NoStop}%
\bibitem [{\citenamefont {Shang}\ \emph {et~al.}(2016)\citenamefont {Shang},
  \citenamefont {Cui}, \citenamefont {Cao}, \citenamefont {Wang}, \citenamefont
  {Chao}, \citenamefont {Shu},\ and\ \citenamefont {Huang}}]{Shang2016}%
  \BibitemOpen
  \bibfield  {author} {\bibinfo {author} {\bibfnamefont {J.-J.}\ \bibnamefont
  {Shang}}, \bibinfo {author} {\bibfnamefont {K.-F.}\ \bibnamefont {Cui}},
  \bibinfo {author} {\bibfnamefont {J.}~\bibnamefont {Cao}}, \bibinfo {author}
  {\bibfnamefont {S.-M.}\ \bibnamefont {Wang}}, \bibinfo {author}
  {\bibfnamefont {S.-J.}\ \bibnamefont {Chao}}, \bibinfo {author}
  {\bibfnamefont {H.-L.}\ \bibnamefont {Shu}},\ and\ \bibinfo {author}
  {\bibfnamefont {X.-R.}\ \bibnamefont {Huang}},\ }\href
  {https://doi.org/10.1088/0256-307X/33/10/103701} {\bibfield  {journal}
  {\bibinfo  {journal} {Chinese Physics Letters}\ }\textbf {\bibinfo {volume}
  {33}},\ \bibinfo {pages} {103701} (\bibinfo {year} {2016})}\BibitemShut
  {NoStop}%
\bibitem [{\citenamefont {Batteiger}\ \emph {et~al.}(2009)\citenamefont
  {Batteiger}, \citenamefont {Kn{\"{u}}nz}, \citenamefont {Herrmann},
  \citenamefont {Saathoff}, \citenamefont {Sch{\"{u}}ssler}, \citenamefont
  {Bernhardt}, \citenamefont {Wilken}, \citenamefont {Holzwarth}, \citenamefont
  {H{\"{a}}nsch},\ and\ \citenamefont {Udem}}]{Batteiger2009}%
  \BibitemOpen
  \bibfield  {author} {\bibinfo {author} {\bibfnamefont {V.}~\bibnamefont
  {Batteiger}}, \bibinfo {author} {\bibfnamefont {S.}~\bibnamefont
  {Kn{\"{u}}nz}}, \bibinfo {author} {\bibfnamefont {M.}~\bibnamefont
  {Herrmann}}, \bibinfo {author} {\bibfnamefont {G.}~\bibnamefont {Saathoff}},
  \bibinfo {author} {\bibfnamefont {H.~A.}\ \bibnamefont {Sch{\"{u}}ssler}},
  \bibinfo {author} {\bibfnamefont {B.}~\bibnamefont {Bernhardt}}, \bibinfo
  {author} {\bibfnamefont {T.}~\bibnamefont {Wilken}}, \bibinfo {author}
  {\bibfnamefont {R.}~\bibnamefont {Holzwarth}}, \bibinfo {author}
  {\bibfnamefont {T.~W.}\ \bibnamefont {H{\"{a}}nsch}},\ and\ \bibinfo {author}
  {\bibfnamefont {T.}~\bibnamefont {Udem}},\ }\href
  {https://doi.org/10.1103/PHYSREVA.80.022503/FIGURES/5/MEDIUM} {\bibfield
  {journal} {\bibinfo  {journal} {Physical Review A - Atomic, Molecular, and
  Optical Physics}\ }\textbf {\bibinfo {volume} {80}},\ \bibinfo {pages}
  {022503} (\bibinfo {year} {2009})}\BibitemShut {NoStop}%
\bibitem [{\citenamefont {Herrmann}\ \emph {et~al.}(2009)\citenamefont
  {Herrmann}, \citenamefont {Batteiger}, \citenamefont {Kn{\"{u}}nz},
  \citenamefont {Saathoff}, \citenamefont {Udem},\ and\ \citenamefont
  {H{\"{a}}nsch}}]{Hermann2009}%
  \BibitemOpen
  \bibfield  {author} {\bibinfo {author} {\bibfnamefont {M.}~\bibnamefont
  {Herrmann}}, \bibinfo {author} {\bibfnamefont {V.}~\bibnamefont {Batteiger}},
  \bibinfo {author} {\bibfnamefont {S.}~\bibnamefont {Kn{\"{u}}nz}}, \bibinfo
  {author} {\bibfnamefont {G.}~\bibnamefont {Saathoff}}, \bibinfo {author}
  {\bibfnamefont {T.}~\bibnamefont {Udem}},\ and\ \bibinfo {author}
  {\bibfnamefont {T.~W.}\ \bibnamefont {H{\"{a}}nsch}},\ }\href
  {https://doi.org/10.1103/PhysRevLettHYSREVLETT.102.013006/FIGURES/5/MEDIUM}
  {\bibfield  {journal} {\bibinfo  {journal} {Physical Review Letters}\
  }\textbf {\bibinfo {volume} {102}},\ \bibinfo {pages} {013006} (\bibinfo
  {year} {2009})}\BibitemShut {NoStop}%
\bibitem [{\citenamefont {Barrett}\ \emph {et~al.}(2003)\citenamefont
  {Barrett}, \citenamefont {DeMarco}, \citenamefont {Schaetz}, \citenamefont
  {Meyer}, \citenamefont {Leibfried}, \citenamefont {Britton}, \citenamefont
  {Chiaverini}, \citenamefont {Itano}, \citenamefont {Jelenkovic},
  \citenamefont {Jost}, \citenamefont {Langer}, \citenamefont {Rosenband},\
  and\ \citenamefont {Wineland}}]{Barrett2003}%
  \BibitemOpen
  \bibfield  {author} {\bibinfo {author} {\bibfnamefont {M.~D.}\ \bibnamefont
  {Barrett}}, \bibinfo {author} {\bibfnamefont {B.}~\bibnamefont {DeMarco}},
  \bibinfo {author} {\bibfnamefont {T.}~\bibnamefont {Schaetz}}, \bibinfo
  {author} {\bibfnamefont {V.}~\bibnamefont {Meyer}}, \bibinfo {author}
  {\bibfnamefont {D.}~\bibnamefont {Leibfried}}, \bibinfo {author}
  {\bibfnamefont {J.}~\bibnamefont {Britton}}, \bibinfo {author} {\bibfnamefont
  {J.}~\bibnamefont {Chiaverini}}, \bibinfo {author} {\bibfnamefont {W.~M.}\
  \bibnamefont {Itano}}, \bibinfo {author} {\bibfnamefont {B.}~\bibnamefont
  {Jelenkovic}}, \bibinfo {author} {\bibfnamefont {J.~D.}\ \bibnamefont
  {Jost}}, \bibinfo {author} {\bibfnamefont {C.}~\bibnamefont {Langer}},
  \bibinfo {author} {\bibfnamefont {T.}~\bibnamefont {Rosenband}},\ and\
  \bibinfo {author} {\bibfnamefont {D.~J.}\ \bibnamefont {Wineland}},\ }\href
  {https://doi.org/10.1103/PhysRevA.68.042302} {\bibfield  {journal} {\bibinfo
  {journal} {Phys. Rev. A}\ }\textbf {\bibinfo {volume} {68}},\ \bibinfo
  {pages} {42302} (\bibinfo {year} {2003})}\BibitemShut {NoStop}%
\bibitem [{\citenamefont {Schwarz}\ \emph {et~al.}(2012)\citenamefont
  {Schwarz}, \citenamefont {Versolato}, \citenamefont {Windberger},
  \citenamefont {Brunner}, \citenamefont {Ballance}, \citenamefont {Eberle},
  \citenamefont {Ullrich}, \citenamefont {Schmidt}, \citenamefont {Hansen},
  \citenamefont {Gingell}, \citenamefont {Drewsen},\ and\ \citenamefont
  {L{\'{o}}pez-Urrutia}}]{Schwarz2012}%
  \BibitemOpen
  \bibfield  {author} {\bibinfo {author} {\bibfnamefont {M.}~\bibnamefont
  {Schwarz}}, \bibinfo {author} {\bibfnamefont {O.~O.}\ \bibnamefont
  {Versolato}}, \bibinfo {author} {\bibfnamefont {A.}~\bibnamefont
  {Windberger}}, \bibinfo {author} {\bibfnamefont {F.~R.}\ \bibnamefont
  {Brunner}}, \bibinfo {author} {\bibfnamefont {T.}~\bibnamefont {Ballance}},
  \bibinfo {author} {\bibfnamefont {S.~N.}\ \bibnamefont {Eberle}}, \bibinfo
  {author} {\bibfnamefont {J.}~\bibnamefont {Ullrich}}, \bibinfo {author}
  {\bibfnamefont {P.~O.}\ \bibnamefont {Schmidt}}, \bibinfo {author}
  {\bibfnamefont {A.~K.}\ \bibnamefont {Hansen}}, \bibinfo {author}
  {\bibfnamefont {A.~D.}\ \bibnamefont {Gingell}}, \bibinfo {author}
  {\bibfnamefont {M.}~\bibnamefont {Drewsen}},\ and\ \bibinfo {author}
  {\bibfnamefont {J.~R.}\ \bibnamefont {L{\'{o}}pez-Urrutia}},\ }\bibfield
  {journal} {\bibinfo  {journal} {Review of Scientific Instruments}\ }\textbf
  {\bibinfo {volume} {83}},\ \href {https://doi.org/10.1063/1.4742770}
  {10.1063/1.4742770} (\bibinfo {year} {2012})\BibitemShut {NoStop}%
\bibitem [{\citenamefont {Maxwell}(1860)}]{Maxwell1860}%
  \BibitemOpen
  \bibfield  {author} {\bibinfo {author} {\bibfnamefont {J.~C.}\ \bibnamefont
  {Maxwell}},\ }\href {https://doi.org/10.1080/14786446008642818} {\bibfield
  {journal} {\bibinfo  {journal} {The London, Edinburgh, and Dublin
  Philosophical Magazine and Journal of Science}\ }\textbf {\bibinfo {volume}
  {19}},\ \bibinfo {pages} {19} (\bibinfo {year} {1860})}\BibitemShut {NoStop}%
\bibitem [{\citenamefont {Kutta}(1901)}]{Kutta1901}%
  \BibitemOpen
  \bibfield  {author} {\bibinfo {author} {\bibfnamefont {W.}~\bibnamefont
  {Kutta}},\ }\href@noop {} {\bibfield  {journal} {\bibinfo  {journal} {Zeit.
  Math. Phys.}\ }\textbf {\bibinfo {volume} {46}},\ \bibinfo {pages} {435}
  (\bibinfo {year} {1901})}\BibitemShut {NoStop}%
\bibitem [{\citenamefont {Drakoudis}\ \emph {et~al.}(2006)\citenamefont
  {Drakoudis}, \citenamefont {S{\"{o}}llner},\ and\ \citenamefont
  {Werth}}]{Drakoudis2006}%
  \BibitemOpen
  \bibfield  {author} {\bibinfo {author} {\bibfnamefont {A.}~\bibnamefont
  {Drakoudis}}, \bibinfo {author} {\bibfnamefont {M.}~\bibnamefont
  {S{\"{o}}llner}},\ and\ \bibinfo {author} {\bibfnamefont {G.}~\bibnamefont
  {Werth}},\ }\href {https://doi.org/10.1016/j.ijms.2006.02.006} {\bibfield
  {journal} {\bibinfo  {journal} {International Journal of Mass Spectrometry}\
  }\textbf {\bibinfo {volume} {252}},\ \bibinfo {pages} {61} (\bibinfo {year}
  {2006})}\BibitemShut {NoStop}%
\bibitem [{\citenamefont {Alheit}\ \emph {et~al.}(1995)\citenamefont {Alheit},
  \citenamefont {Hennig}, \citenamefont {Morgenstern}, \citenamefont {Vedel},\
  and\ \citenamefont {Werth}}]{Alheit1995}%
  \BibitemOpen
  \bibfield  {author} {\bibinfo {author} {\bibfnamefont {R.}~\bibnamefont
  {Alheit}}, \bibinfo {author} {\bibfnamefont {C.}~\bibnamefont {Hennig}},
  \bibinfo {author} {\bibfnamefont {R.}~\bibnamefont {Morgenstern}}, \bibinfo
  {author} {\bibfnamefont {F.}~\bibnamefont {Vedel}},\ and\ \bibinfo {author}
  {\bibfnamefont {G.}~\bibnamefont {Werth}},\ }\href
  {https://doi.org/10.1007/BF01082047/METRICS} {\bibfield  {journal} {\bibinfo
  {journal} {Applied Physics B Lasers and Optics}\ }\textbf {\bibinfo {volume}
  {61}},\ \bibinfo {pages} {277} (\bibinfo {year} {1995})}\BibitemShut
  {NoStop}%
\bibitem [{\citenamefont {Eades}\ \emph {et~al.}(1992)\citenamefont {Eades},
  \citenamefont {Yost},\ and\ \citenamefont {Cooks}}]{Eades1992}%
  \BibitemOpen
  \bibfield  {author} {\bibinfo {author} {\bibfnamefont {D.~M.}\ \bibnamefont
  {Eades}}, \bibinfo {author} {\bibfnamefont {R.~A.}\ \bibnamefont {Yost}},\
  and\ \bibinfo {author} {\bibfnamefont {R.~G.}\ \bibnamefont {Cooks}},\ }\href
  {https://doi.org/10.1002/RCM.1290060908} {\bibfield  {journal} {\bibinfo
  {journal} {Rapid Communications in Mass Spectrometry}\ }\textbf {\bibinfo
  {volume} {6}},\ \bibinfo {pages} {573} (\bibinfo {year} {1992})}\BibitemShut
  {NoStop}%
\bibitem [{\citenamefont {Harmon}\ \emph {et~al.}(2003)\citenamefont {Harmon},
  \citenamefont {Moazzan-Ahmadi},\ and\ \citenamefont {Thompson}}]{Harmon2003}%
  \BibitemOpen
  \bibfield  {author} {\bibinfo {author} {\bibfnamefont {T.~J.}\ \bibnamefont
  {Harmon}}, \bibinfo {author} {\bibfnamefont {N.}~\bibnamefont
  {Moazzan-Ahmadi}},\ and\ \bibinfo {author} {\bibfnamefont {R.~I.}\
  \bibnamefont {Thompson}},\ }\href
  {https://doi.org/10.1103/PhysRevA.67.013415} {\bibfield  {journal} {\bibinfo
  {journal} {Physical Review A}\ }\textbf {\bibinfo {volume} {67}},\ \bibinfo
  {pages} {013415} (\bibinfo {year} {2003})}\BibitemShut {NoStop}%
\bibitem [{\citenamefont {Harris}\ \emph {et~al.}(2020)\citenamefont {Harris},
  \citenamefont {Millman}, \citenamefont {van~der Walt}, \citenamefont
  {Gommers}, \citenamefont {Virtanen}, \citenamefont {Cournapeau},
  \citenamefont {Wieser}, \citenamefont {Taylor}, \citenamefont {Berg},
  \citenamefont {Smith}, \citenamefont {Kern}, \citenamefont {Picus},
  \citenamefont {Hoyer}, \citenamefont {van Kerkwijk}, \citenamefont {Brett},
  \citenamefont {Haldane}, \citenamefont {del R{\'{i}}o}, \citenamefont
  {Wiebe}, \citenamefont {Peterson}, \citenamefont {G{\'{e}}rard-Marchant},
  \citenamefont {Sheppard}, \citenamefont {Reddy}, \citenamefont {Weckesser},
  \citenamefont {Abbasi}, \citenamefont {Gohlke},\ and\ \citenamefont
  {Oliphant}}]{harris2020array}%
  \BibitemOpen
  \bibfield  {author} {\bibinfo {author} {\bibfnamefont {C.~R.}\ \bibnamefont
  {Harris}}, \bibinfo {author} {\bibfnamefont {K.~J.}\ \bibnamefont {Millman}},
  \bibinfo {author} {\bibfnamefont {S.~J.}\ \bibnamefont {van~der Walt}},
  \bibinfo {author} {\bibfnamefont {R.}~\bibnamefont {Gommers}}, \bibinfo
  {author} {\bibfnamefont {P.}~\bibnamefont {Virtanen}}, \bibinfo {author}
  {\bibfnamefont {D.}~\bibnamefont {Cournapeau}}, \bibinfo {author}
  {\bibfnamefont {E.}~\bibnamefont {Wieser}}, \bibinfo {author} {\bibfnamefont
  {J.}~\bibnamefont {Taylor}}, \bibinfo {author} {\bibfnamefont
  {S.}~\bibnamefont {Berg}}, \bibinfo {author} {\bibfnamefont {N.~J.}\
  \bibnamefont {Smith}}, \bibinfo {author} {\bibfnamefont {R.}~\bibnamefont
  {Kern}}, \bibinfo {author} {\bibfnamefont {M.}~\bibnamefont {Picus}},
  \bibinfo {author} {\bibfnamefont {S.}~\bibnamefont {Hoyer}}, \bibinfo
  {author} {\bibfnamefont {M.~H.}\ \bibnamefont {van Kerkwijk}}, \bibinfo
  {author} {\bibfnamefont {M.}~\bibnamefont {Brett}}, \bibinfo {author}
  {\bibfnamefont {A.}~\bibnamefont {Haldane}}, \bibinfo {author} {\bibfnamefont
  {J.~F.}\ \bibnamefont {del R{\'{i}}o}}, \bibinfo {author} {\bibfnamefont
  {M.}~\bibnamefont {Wiebe}}, \bibinfo {author} {\bibfnamefont
  {P.}~\bibnamefont {Peterson}}, \bibinfo {author} {\bibfnamefont
  {P.}~\bibnamefont {G{\'{e}}rard-Marchant}}, \bibinfo {author} {\bibfnamefont
  {K.}~\bibnamefont {Sheppard}}, \bibinfo {author} {\bibfnamefont
  {T.}~\bibnamefont {Reddy}}, \bibinfo {author} {\bibfnamefont
  {W.}~\bibnamefont {Weckesser}}, \bibinfo {author} {\bibfnamefont
  {H.}~\bibnamefont {Abbasi}}, \bibinfo {author} {\bibfnamefont
  {C.}~\bibnamefont {Gohlke}},\ and\ \bibinfo {author} {\bibfnamefont {T.~E.}\
  \bibnamefont {Oliphant}},\ }\href {https://doi.org/10.1038/s41586-020-2649-2}
  {\bibfield  {journal} {\bibinfo  {journal} {Nature}\ }\textbf {\bibinfo
  {volume} {585}},\ \bibinfo {pages} {357} (\bibinfo {year}
  {2020})}\BibitemShut {NoStop}%
\bibitem [{\citenamefont {Virtanen}\ \emph {et~al.}(2020)\citenamefont
  {Virtanen}, \citenamefont {Gommers}, \citenamefont {Oliphant}, \citenamefont
  {Haberland}, \citenamefont {Reddy}, \citenamefont {Cournapeau}, \citenamefont
  {Burovski}, \citenamefont {Peterson}, \citenamefont {Weckesser},
  \citenamefont {Bright}, \citenamefont {{van der Walt}}, \citenamefont
  {Brett}, \citenamefont {Wilson}, \citenamefont {Millman}, \citenamefont
  {Mayorov}, \citenamefont {Nelson}, \citenamefont {Jones}, \citenamefont
  {Kern}, \citenamefont {Larson}, \citenamefont {Carey}, \citenamefont {Polat},
  \citenamefont {Feng}, \citenamefont {Moore}, \citenamefont {{VanderPlas}},
  \citenamefont {Laxalde}, \citenamefont {Perktold}, \citenamefont {Cimrman},
  \citenamefont {Henriksen}, \citenamefont {Quintero}, \citenamefont {Harris},
  \citenamefont {Archibald}, \citenamefont {Ribeiro}, \citenamefont
  {Pedregosa}, \citenamefont {{van Mulbregt}},\ and\ \citenamefont {{SciPy 1.0
  Contributors}}}]{2020SciPy-NMeth}%
  \BibitemOpen
  \bibfield  {author} {\bibinfo {author} {\bibfnamefont {P.}~\bibnamefont
  {Virtanen}}, \bibinfo {author} {\bibfnamefont {R.}~\bibnamefont {Gommers}},
  \bibinfo {author} {\bibfnamefont {T.~E.}\ \bibnamefont {Oliphant}}, \bibinfo
  {author} {\bibfnamefont {M.}~\bibnamefont {Haberland}}, \bibinfo {author}
  {\bibfnamefont {T.}~\bibnamefont {Reddy}}, \bibinfo {author} {\bibfnamefont
  {D.}~\bibnamefont {Cournapeau}}, \bibinfo {author} {\bibfnamefont
  {E.}~\bibnamefont {Burovski}}, \bibinfo {author} {\bibfnamefont
  {P.}~\bibnamefont {Peterson}}, \bibinfo {author} {\bibfnamefont
  {W.}~\bibnamefont {Weckesser}}, \bibinfo {author} {\bibfnamefont
  {J.}~\bibnamefont {Bright}}, \bibinfo {author} {\bibfnamefont {S.~J.}\
  \bibnamefont {{van der Walt}}}, \bibinfo {author} {\bibfnamefont
  {M.}~\bibnamefont {Brett}}, \bibinfo {author} {\bibfnamefont
  {J.}~\bibnamefont {Wilson}}, \bibinfo {author} {\bibfnamefont {K.~J.}\
  \bibnamefont {Millman}}, \bibinfo {author} {\bibfnamefont {N.}~\bibnamefont
  {Mayorov}}, \bibinfo {author} {\bibfnamefont {A.~R.~J.}\ \bibnamefont
  {Nelson}}, \bibinfo {author} {\bibfnamefont {E.}~\bibnamefont {Jones}},
  \bibinfo {author} {\bibfnamefont {R.}~\bibnamefont {Kern}}, \bibinfo {author}
  {\bibfnamefont {E.}~\bibnamefont {Larson}}, \bibinfo {author} {\bibfnamefont
  {C.~J.}\ \bibnamefont {Carey}}, \bibinfo {author} {\bibfnamefont
  {{\.I}.}~\bibnamefont {Polat}}, \bibinfo {author} {\bibfnamefont
  {Y.}~\bibnamefont {Feng}}, \bibinfo {author} {\bibfnamefont {E.~W.}\
  \bibnamefont {Moore}}, \bibinfo {author} {\bibfnamefont {J.}~\bibnamefont
  {{VanderPlas}}}, \bibinfo {author} {\bibfnamefont {D.}~\bibnamefont
  {Laxalde}}, \bibinfo {author} {\bibfnamefont {J.}~\bibnamefont {Perktold}},
  \bibinfo {author} {\bibfnamefont {R.}~\bibnamefont {Cimrman}}, \bibinfo
  {author} {\bibfnamefont {I.}~\bibnamefont {Henriksen}}, \bibinfo {author}
  {\bibfnamefont {E.~A.}\ \bibnamefont {Quintero}}, \bibinfo {author}
  {\bibfnamefont {C.~R.}\ \bibnamefont {Harris}}, \bibinfo {author}
  {\bibfnamefont {A.~M.}\ \bibnamefont {Archibald}}, \bibinfo {author}
  {\bibfnamefont {A.~H.}\ \bibnamefont {Ribeiro}}, \bibinfo {author}
  {\bibfnamefont {F.}~\bibnamefont {Pedregosa}}, \bibinfo {author}
  {\bibfnamefont {P.}~\bibnamefont {{van Mulbregt}}},\ and\ \bibinfo {author}
  {\bibnamefont {{SciPy 1.0 Contributors}}},\ }\href
  {https://doi.org/10.1038/s41592-019-0686-2} {\bibfield  {journal} {\bibinfo
  {journal} {Nature Methods}\ }\textbf {\bibinfo {volume} {17}},\ \bibinfo
  {pages} {261} (\bibinfo {year} {2020})}\BibitemShut {NoStop}%
\end{thebibliography}%
%%%%%%%%%%%%%%%%%%%%%%%%%%%%%%%%%%%%%%%%%%%%%
\end{document}